\documentclass[twocolumn,english,groupedaddress, superscriptaddress,aps,pre,10pt,floatfix]{revtex4-2}
\usepackage[T1]{fontenc}
\usepackage[utf8]{inputenc}
\usepackage{amsmath,amsthm,mathtools,amssymb}
\usepackage{graphicx}
\usepackage{siunitx}
\usepackage{orcidlink}
\usepackage{svg}
\usepackage{xurl}

\begin{document}

\title{Cyber-Physical Systems on the Megawatt Scale: The impact of battery control on grid frequency stability.}

\author{Carsten Hartmann \orcidlink{0009-0007-5067-589X} }
\affiliation{Institute of Climate and Energy Systems: Energy Systems Engineering (ICE-1), Forschungszentrum J\"ulich, 52428 J\"ulich, Germany}
\affiliation{RWTH Aachen University, 52062 Aachen, Germany}

\author{Edoardo De Din \orcidlink{0000-0003-2271-4808}}
\affiliation{Institute of Climate and Energy Systems: Energy Systems Engineering (ICE-1), Forschungszentrum J\"ulich, 52428 J\"ulich, Germany}

\author{Daniele Carta \orcidlink{0000-0002-0182-8710}}
\affiliation{Institute of Climate and Energy Systems: Energy Systems Engineering (ICE-1), Forschungszentrum J\"ulich, 52428 J\"ulich, Germany}

\author{Florian Middelkoop}
\affiliation{Institute for Theoretical Physics, University of Cologne, 50937 K\"oln, Germany}

\author{Arndt Neubauer}
\affiliation{TenneT TSO GmbH, Bernecker Str. 70, 95448 Bayreuth, Germany}

\author{Johannes Kruse \orcidlink{0000-0002-3478-3379}}
\affiliation{Institute of Climate and Energy Systems: Energy Systems Engineering (ICE-1), Forschungszentrum J\"ulich, 52428 J\"ulich, Germany}
\affiliation{Institute for Theoretical Physics, University of Cologne, 50937 K\"oln, Germany}

\author{Ulrich Oberhofer \orcidlink{0009-0001-2766-5796}}
\affiliation{Institute for Automation and Applied Informatics, Karlsruhe Institute of Technology, 76344 Eggenstein-Leopoldshafen, Germany}

\author{Richard Jumar \orcidlink{0000-0001-6854-4678}}
\affiliation{Institute for Automation and Applied Informatics, Karlsruhe Institute of Technology, 76344 Eggenstein-Leopoldshafen, Germany}

\author{Benjamin Schäfer \orcidlink{0000-0003-1607-9748}}
\affiliation{Institute for Automation and Applied Informatics, Karlsruhe Institute of Technology, 76344 Eggenstein-Leopoldshafen, Germany}

\author{Thiemo Pesch \orcidlink{0000-0002-3297-6599}}
\affiliation{Institute of Climate and Energy Systems: Energy Systems Engineering (ICE-1), Forschungszentrum J\"ulich, 52428 J\"ulich, Germany}

\author{Andrea Benigni \orcidlink{0000-0002-2475-7003}}
\affiliation{Institute of Climate and Energy Systems: Energy Systems Engineering (ICE-1), Forschungszentrum J\"ulich, 52428 J\"ulich, Germany}
\affiliation{RWTH Aachen University, 52062 Aachen, Germany}
\affiliation{JARA-ENERGY, 52074 Aachen, Germany}

\author{Dirk Witthaut \orcidlink{0000-0002-3623-5341}}
\affiliation{Institute of Climate and Energy Systems: Energy Systems Engineering (ICE-1), Forschungszentrum J\"ulich, 52428 J\"ulich, Germany}
\affiliation{Institute for Theoretical Physics, University of Cologne, 50937 K\"oln, Germany}
\affiliation{JARA-ENERGY, 52074 Aachen, Germany}

\begin{abstract}
Electric power systems are undergoing fundamental change. The shift to inverter-based generation challenges frequency stability, while growing digitalisation heightens vulnerability to errors and attacks. Here we identify an emerging risk at the intersection of cyber–physical coupling and control system design. We show that grid frequency time series worldwide exhibit a persistent one-minute oscillatory pattern, whose origin has remained largely unexplained.  We trace this pattern back to the energy management systems of battery electric storage systems and  demonstrate that the pattern amplitude has increased substantially in the Nordic and British grids. We argue that this effect is a potential burden for stability in future grids with low inertia and an increasing penetration with batteries and smart devices, though it can be mitigated by a revision of battery control algorithms.
\end{abstract}

\maketitle

\section{Introduction} 

The transition to a sustainable energy system requires advanced control methods to deal with increasing complexity and uncertainty~\cite{denholm2021challenges,marot2022perspectives}. As variable renewable energy sources like wind and solar replace conventional generation, maintaining balance requires accurate forecasting and real-time control of storage, backup generation, and flexible demand~\cite{morales2013integrating}. At the same time, the electrification of sectors such as transport and heating is introducing millions of digitally controlled devices -- electric vehicle chargers~\cite{muratori2018impact}, heat pumps~\cite{liu2024short}, and smart appliances~\cite{farhangi2009path} -- into the grid. Together, these trends are reshaping the power system into a cyber-physical infrastructure, where physical devices and digital control are deeply intertwined. In this new paradigm, system reliability increasingly hinges on the quality and coordination of control algorithms across all layers of operation~\cite{xie2021toward}.

While cyber-physical integration enables unprecedented levels of efficiency and flexibility, it also introduces new systemic risks~\cite{pasqualetti2013attack}. Historical events underscore the stakes: inadequate situational awareness contributed to the 2003 North American blackout~\cite{andersson2005causes}; a coordinated cyber attack caused widespread outages in Ukraine in 2016~\cite{liang20162015}; and inadequate failure response by wind turbine controllers played a critical role in the 2016 South Australian blackout~\cite{yan2018anatomy}. 

In this article, we focus on another, less conspicuous but equally pervasive burden for grid stability -- one that manifests daily in power systems around the world. We show that the grid frequency and thus the power balance exhibit a periodic pattern with a period of exactly 1 minute. We trace this pattern back to the controllers, i.e.~the energy management system (EMS), of uninterruptible power supplies (UPS) based on battery electric storage systems (BESS). We further show how this pattern evolved over time and discuss its potential impact on the stability of future grids.

Some aspects of low-frequency oscillations have been discussed before, but the root causes remained elusive. A report by Nordic transmission system operators suggests that water turbine systems can amplify oscillations of the power imbalance with a period around $1 \, \si{min}$~\cite{oftebro2014analysis,odelbrink2014design,fcp_project_summary_2017}. A recent study~\cite{wei2025freqpats} speculates that the pattern may be partly rooted in the secondary control system. 

\begin{figure*}[tb]
    \centering
    \includegraphics[width=\textwidth]{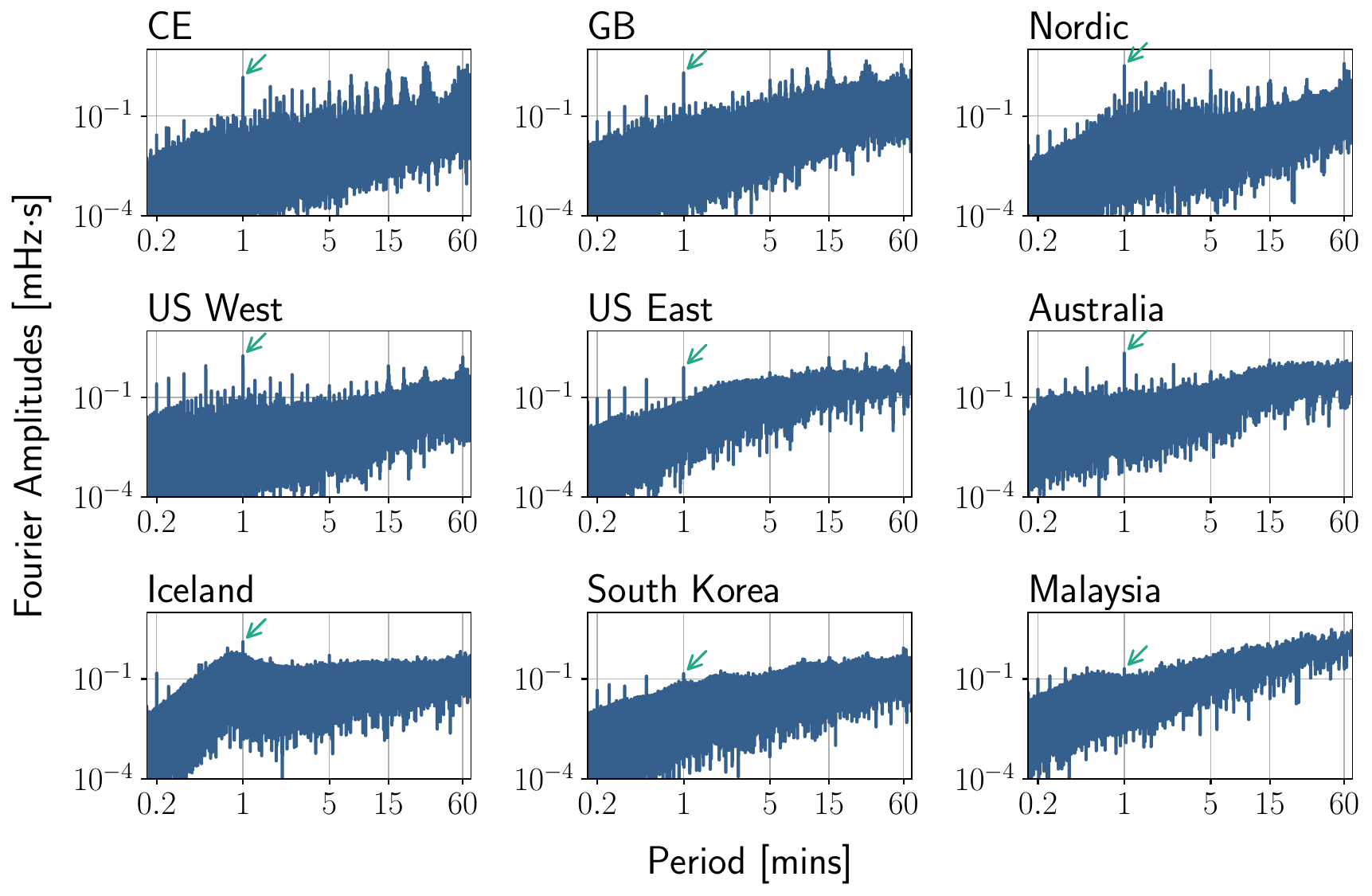}
    \caption{Fourier Spectra of the grid frequency $f(t)$ for nine grids on different continents that exhibit a $1 \, \si{min}$ pattern. Measurements for the CE, GB and Nordic grids are taken on the transmission grid level, for the other grids on the distribution grid level. Details on the data sources and processing are described in the Methods section.
    }
    \label{fig:fourier_freq_grids}
\end{figure*}

BESS provides one example of devices operating with algorithmically synchronized or clocked control algorithms. With the ongoing ``smartification'' of electrical equipment, an increasing number of devices can be expected to exhibit similar behaviour. At the same time, grids are becoming more susceptible to power imbalances due to the decreasing inertia~\cite{Milano2018FoundationsAC}. Indeed, we observe an increasing pattern amplitude in certain grids. Our results highlight the crucial role of adequate control algorithms for the stability of future power grids.

\section{A universal $1 \, \si{min}$ pattern in the grid frequency}

The grid frequency is the key observable in power system control, reflecting the balance between generation and load: it rises when generation exceeds demand and falls when demand outpaces supply. To maintain a nominal reference of $50 \, \si{Hz}$ or $60 \, \si{Hz}$, depending on the region, sophisticated control systems continuously adjust generation~\cite{handbook2009policy}.

Large deviations from this reference occur during contingencies and can endanger system stability, prompting stringent security protocols~\cite{de_boeck_review_2016}. For example, the recent Iberian blackout saw the local frequency drop to $48 \si{Hz}$ prior to the outage~\cite{entsoe2025iberian}. Smaller, more frequent deviations arise from both stochastic fluctuations and deterministic mechanisms~\cite{schaferNonGaussianPowerGrid2018,entsoe_dfd2020}. 

Recurrent patterns are found on different time scales. We carry out a Fourier analysis of the grid frequency time series $f(t)$ for a vary of power grids to identify the dominant contributions (Fig.~\ref{fig:fourier_freq_grids} and \ref{fig:fourier_no_peaks_freq}). The selection of grids covers three continents, with a size spanning from approximately 400,000 customers in Iceland to more than 400 million customers in the Central European (CE) grid.

For the European grids, we observe pronounced peaks at a period of $15 \, \si{min}$ and $60 \, \si{min}$ and their higher harmonics. These recurrent patterns can be attributed to the impact of electricity trading and are well known in the literature ~\cite{weissbachHighFrequencyDeviations2009,kruse2021exploring}. Electricity is traded in fixed blocks of $15 \si{min}$ and $60 \, \si{min}$, leading to abrupt changes in the scheduled generation at the beginning of each interval.

In addition, there is a recurrent variation of the grid frequency with a period of 1 minute. We observe peaks in the Fourier spectrum at $1 \, \si{min}$ and higher harmonics for various power grids on four different continents (Fig.~\ref{fig:fourier_freq_grids}). The $1 \, \si{min}$ peak is particularly pronounced in large interconnected grids in Europe, North America, and Australia, but also visible in Iceland, South Korea and Malaysia. The peak is extremely narrow, pointing to a pattern that has a precisely defined frequency and is stable over years. What causes this universal pattern?

\section{Characterization of the $1 \, \si{min}$ pattern.}

\begin{figure*}
    \centering
    \includegraphics[width=\textwidth]{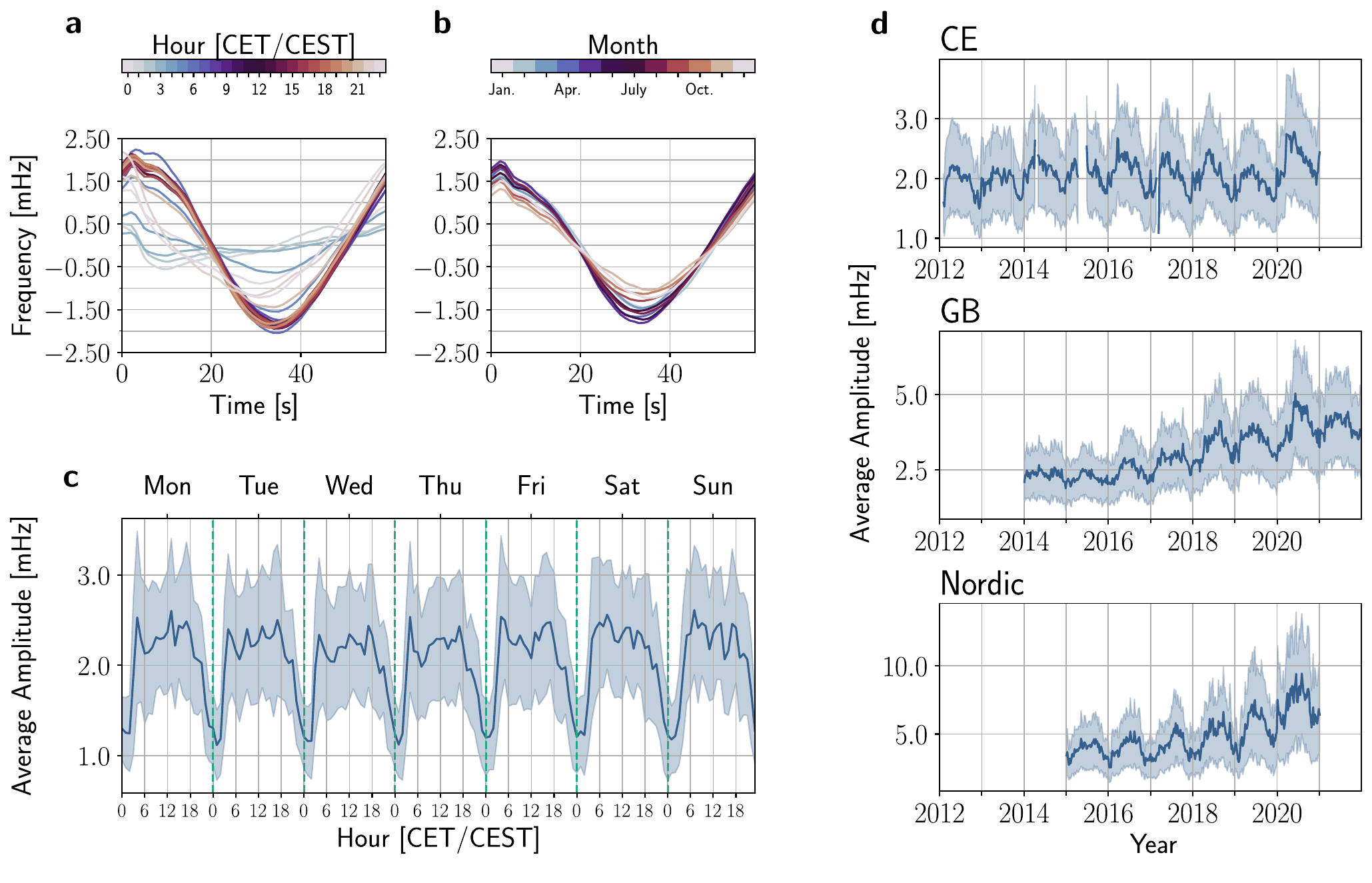}
    \caption{Characterization of the $1 \, \si{min}$ pattern.
    \textbf{a} Average Waveform over a day in local time, that is Central European Time (CET)  and Central European Summer Time (CEST) for the CE grid in 2018. The pattern's amplitude changes during the day, but the overall shape remains consistent: The frequency consistently decreases during the first half of the minute. 
    \textbf{b} Average Waveform over the year for the CE grid. The pattern is highly stable; only a slight seasonal profile in the amplitudes is visible. \textbf{c} Weekly profile of the amplitudes for the CE grid in 2018. The pattern is significantly weaker during the night. On weekends, the morning increase is later than on weekdays. Time is given in local time, and the shaded area corresponds to the standard deviation. 
    \textbf{d} Long-term trends for the weekly averaged amplitudes for the three major European grids. We observe a seasonal pattern: the amplitudes are stronger in summer than in winter, except for the Christmas/New Year period. For the GB and Nordic grids, the strength is also clearly increasing over the years. During the COVID-19 lockdowns in spring 2020, we observed stronger amplitudes for all grids. Again, the shaded area corresponds to the standard deviation. 
    }
    \label{fig:signal_char_CE}
\end{figure*}

In order to characterize the universal pattern, we divide the frequency time series $f(t)$ into chunks of length $T$. For every chunk, we compute a characteristic minutely profile $s(t)$ that is referred to as wave form in the following. More precisely, $s(t)$ is the average over all minutes in the respective chunk for a given second of the minute $t$ (see Methods for details). Typically we use chunks of length $T = 1 \ \mathrm{hour}$ to remove the hourly patterns induced by market effects, indicated by the subscript $h$. Furthermore, we define the hourly amplitude as $a_h = \max_t(s_h(t)) - \min_t(s_h(t))$ to characterise the strength of the pattern.

We show averaged results for the CE grid in Fig.~\ref{fig:signal_char_CE}a-c. We observe that the pattern is substantially stronger during the day than during the night. The amplitude reaches a minimum shortly after midnight and remains high during the day. In contrast, there is no pronounced variation between the month of the year or the days of the week. One aspect holds for all hours of the day: The pattern $s_h(t)$ is high at the beginning of the minute and then decreases for few seconds (night) up to  half a minute (day). Comparable effects are observed for other power grids: Typically, the pattern $s_h(t)$ decreases in the beginning or during the first half of the minute (cf.~Fig.~\ref{fig:waveforms_freq_grids}). For instance, we observe a rather sharp drop in the beginning of the minute followed by a slower recovery.

The stability of the observed pattern is remarkable. We find that the overall shape does not drift, but stays stable over the years. Hence, it appears unlikely that the pattern is due to an intrinsic oscillatory mode. Rather, it is most likely that the drivers are synchronized to a stable clock.

These findings point to a universal pattern in the power balance. In general, the rate of change of frequency (RoCoF) is proportional to the power imbalance in the grid (see Methods). Hence, the observed decrease of the frequency indicates that demand exceeds generation in the first half of the minute. This pattern is found in power systems across different continents.

We note, however, that the detailed relation of frequency and power is more involved. The grid inertia smooths the frequency response similar to a low-pass filter, and resonance phenomena may amplify the response at certain frequencies. This includes in particular, resonances in  hydro power turbine systems that may amplify the observed pattern~\cite{oftebro2014analysis,odelbrink2014design,fcp_project_summary_2017}.

\section{Long term trends}

The characteristic shape of the $1 \, \si{min}$ pattern is remarkably stable, while its amplitude is subject to change. We investigate the long-term trend for three European grids where sufficient data is available (Fig.~\ref{fig:signal_char_CE}d). We observe that the amplitude has increased substantially in the British (GB) and the Nordic Grid and mildly in the CE grid. In the Nordic grid, the amplitude regularly exceeds $10 \, \si{mHz}$.

Two factors can contribute to the observed increase. In general, the RoCoF is proportional to the power imbalance divided by the effective inertia of the grid (see Methods). Inertia is mainly provided by synchronous machines and thus decreases during the transition to inverter-based renewable power sources~
\cite{Milano2018FoundationsAC}. This directly leads to an increase of the RoCoF and explains the observed increase of the pattern amplitude.

Furthermore, we observe a temporary increase in the pattern amplitude during the Covid-19 lockdown (see Fig.~\ref{fig:amplitude-lockdown}). It is plausible that the reduction of inertia from industrial synchronous motors contributed to this effect, as industry production decreased substantially during the lockdown. However, this does not rule out an increase of the power imbalance. For instance, the internet traffic increased during the lockdown, which probably led to higher power demands.

In conclusion, the pattern amplitude in the Nordic grid has reached a level that is highly relevant for system operation. Regulations do not foresee a deadband in the Frequency containment reserves (FCR-N) in the Nordic grid~\cite{entsoe2024overview}. Hence, the $1 \, \si{min}$ pattern regularly trigger the activation of control reserves. Furthermore, it adds up to other deterministic or stochastic disturbance and thus contributes substantially to large frequency deviations. As inertia decreases, these effects will become increasingly relevant in the future.

\section{From frequency to power}

\begin{figure}
    \centering
    \includegraphics[width=\columnwidth]{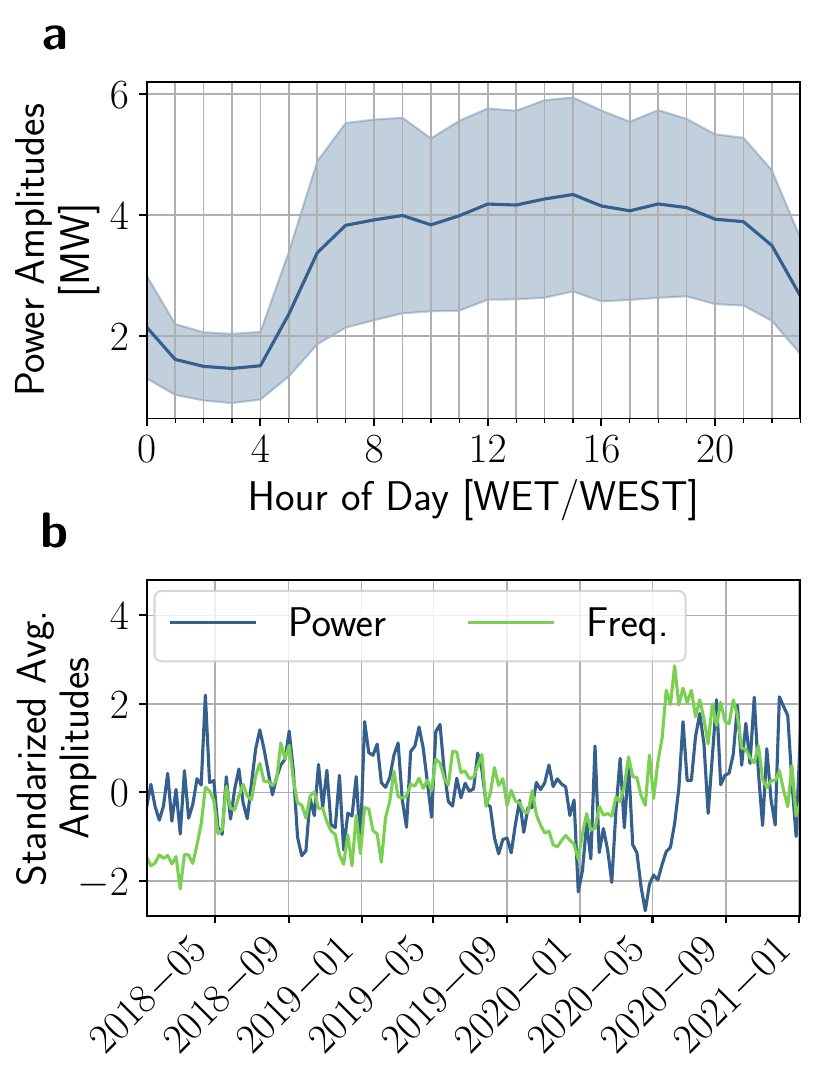}
    \caption{Statistical Analysis of the power amplitudes $P_{T}(h)$, thus removing the impact of inertia, for the GB grid. 
    \textbf{a:} The power amplitudes show a profound daily pattern, similar to the amplitudes in the frequency. The hour is given in local time, and the shaded area corresponds to the standard deviation. 
    \textbf{b:} Comparison of the weekly averaged and standardized power and frequency amplitudes time series. We conclude that the seasonal pattern and the long-term trend in the frequency amplitudes can be attributed to changes in the grid inertia.
    }
    \label{fig:amplitude_power_gb}
\end{figure}

For a further analysis, we have to abstract from the inertia and focus on real power. The British National Energy System Operator (NESO) publishes data on the estimated inertia, such that we can infer the $1 \, \si{min}$ pattern in the active power imbalance (see methods). 

The amplitude of the active power pattern shows a pronounced daily pattern, consistent with the findings for the frequency $1 \, \si{min}$ pattern  (Fig.~\ref{fig:amplitude_power_gb}a). The amplitudes are large during the day and small during the night, with a pronounced increase at the start of the working day. 
In contrast, the power amplitude does not show a seasonal pattern while the frequency amplitude does (Fig.~\ref{fig:amplitude_power_gb}b). Hence, inertia can explain the seasonal variability but not the daily variability of the pattern amplitude.

Investigating the data over longer periods shows that the amplitude of the $1 \, \si{min}$ power pattern does not increase significantly -- in contrast to the amplitude of $1 \, \si{min}$ frequency pattern (Fig.~\ref{fig:amplitude_power_gb}b). Hence, we can attribute the long-term trends to the ongoing decrease in grid inertia as discussed above.

\section{Identifying the drivers of the $1 \, \si{min}$ pattern}

\begin{figure}
    \centering
    \includegraphics[width=\columnwidth]{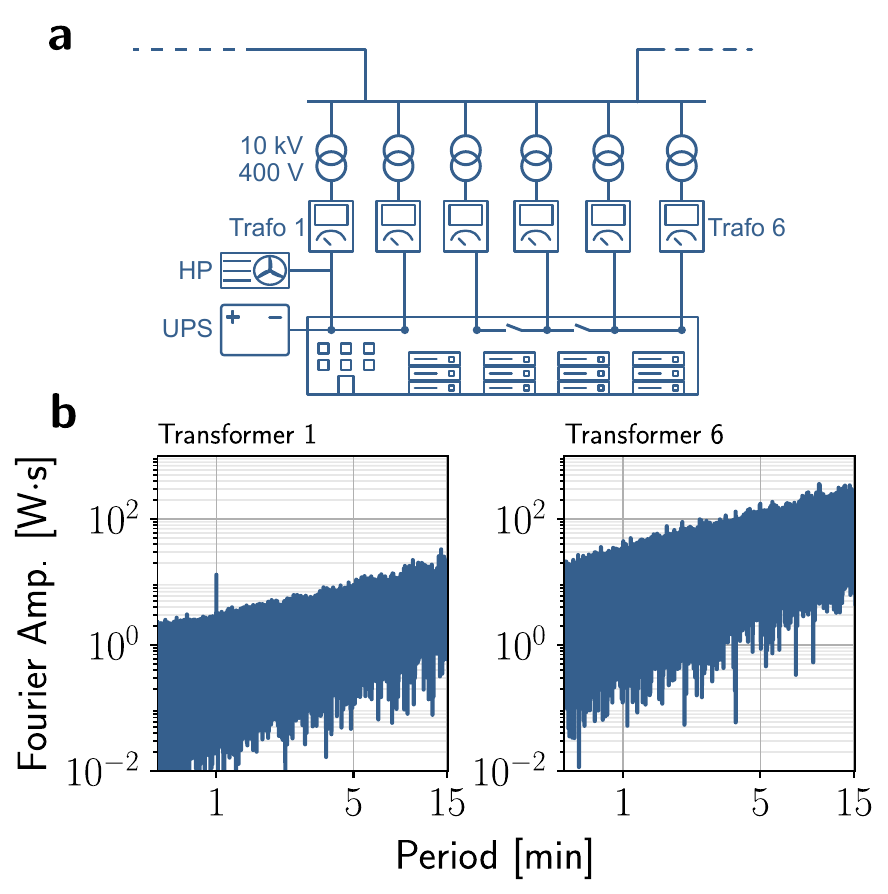}
    \caption{Fourier Spectrum of the Active Power Time Series for selected PQ-Meters in a building on the Jülich research campus. \textbf{a:} Electrical Diagram of Building 1 specifying the location of the PQ Meters at the Transformers. The building contains offices and computing infrastructures. \textbf{b:} For Transformer 1 we measure a $1 \, \si{min}$ pattern in the active power demand. Attached to this transformer are offices, a heat pump, and a UPS. For Transformer 6, we do not observe any significant patterns in the same measurement period. However, the noise level is also significantly higher, and potential patterns could be hidden.  
    }
    \label{fig:fourier_power_campus_16_16}
\end{figure}

Which devices ultimately cause the $1 \, \si{min}$ pattern  in the power imbalance and the grid frequency? 

An examination of the wave form in the CE grid at the distribution level shows that it is slightly phase-advanced compared to the transmission level (cf.~Fig.~\ref{fig:geographic_distribution_2019}). This phase shift suggests that the driving sources are located in the distribution grid, while the large synchronous generators in the transmission system are being driven. Hence, measurements on distribution grids are needed for a detailed analysis.

Forschungszentrum Jülich (FZJ) operates a comprehensive measurement infrastructure in the campus distribution grid. We analyze time series of the real power consumption $P(t)$ from various buildings and transformers to identify possible drivers. 

Remarkably, a $1 \, \si{min}$ component in the active power time series is observed only for some locations, but is not found elsewhere. Two examples are shown in Fig.~\ref{fig:fourier_power_campus_16_16}, and further data is presented in the Fig.~\ref{fig:campus_active_fourier}  and Fig.~\ref{fig:campus_reactive_fourier}.
We find that all measurements showing a $1 \, \si{min}$ component are taken on transformers that serve an uninterruptible power supply (UPS). These UPS are battery-based and provide power to the IT infrastructure and sensitive loads in the event of a grid outage.  

Based on these results, we hypothesize that the EMS of BESSs, in particular UPSs, are the primary drivers of the $1 \, \si{min}$ pattern across the world. The control algorithms of the EMS drive a periodic increase in the power demand every minute. This conclusion is further supported by the fact that the grid frequency on the Spanish island of Mallorca showed a $1 \, \si{min}$ pattern only after the installation of BESS in 2024 (cf.~appendix~\ref{sec:Mallorca}).

\section{The role of battery controllers}

\begin{figure*}
    \centering
    \includegraphics[width=\textwidth, trim=0cm 0cm 0cm 0cm, clip]{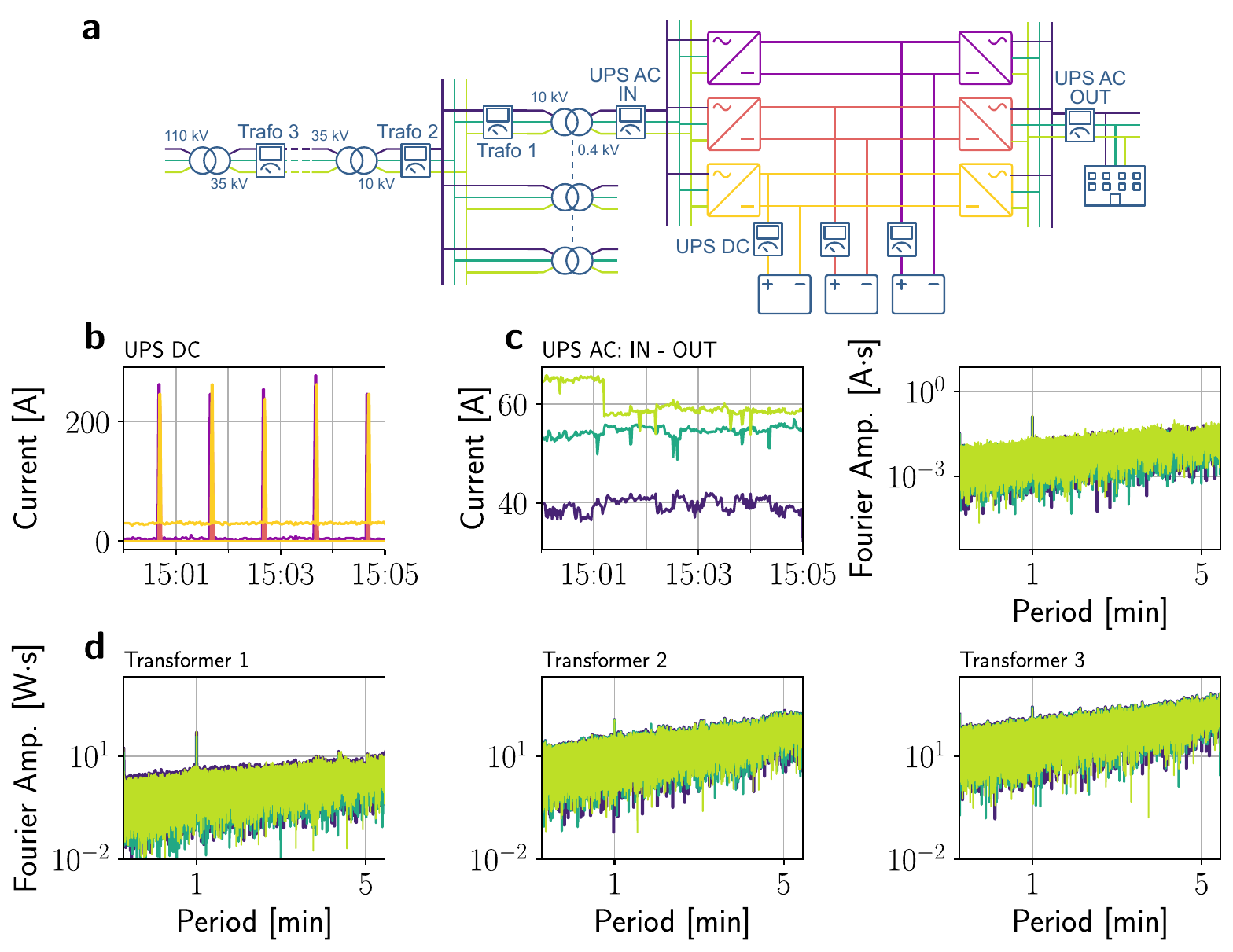}
    \caption{Recurrent Current Peaks in a UPS cause periodic $1 \, \si{min}$power variations across all voltage levels up to the connection to $110 \, \si{kV}$ grid.
    \textbf{a} Electrical diagram of the UPS at Building 3 and its connection to the $110 \, \si{kV}$ distribution grid, indicating the locations of all measurement devices used to trace the pattern. 
    \textbf{b} Recurrent narrow current peaks with $1 \, \si{min}$ periodicity are observed at the DC side of the battery. 
    \textbf{c} 
    Although no clear $1 \, \si{min}$ pattern is evident in the three-phase AC currents immediately before the grid side inverter, Fourier analysis reveals a $1 \, \si{min}$ component. To exclude the influence of downstream loads, the AC output current after the building side inverter is subtracted such that only the internal consumption is considered.
    \textbf{d} Fourier spectra show that the $1 \, \si{min}$ current pattern gives rise to a corresponding pattern in the active power demand, which can be traced across two transformers and a bus bar up to the connection with the $110 \, \si{kV}$ distribution grid.}
    \label{fig:ups_driver}
\end{figure*}

In order to elucidate the role of UPSs in grid power balancing, we are conducting a detailed analysis of one such device located in Building 3 on the Jülich campus (cf.~Table~\ref{tab:pq_meter_data}). 
Comprehensive measurement data is available for this UPS including (i) the DC current to the battery, (ii) the AC currents and power flows before and after the UPS and (iii) the power flow on all transformers linking the UPS to the public $110 \, \si{kV}$ distribution grid, see Fig.~\ref{fig:ups_driver}a for a schematic electrical diagram specifying the measurement locations. 

On the DC side of the UPS, at the battery terminals, narrow current peaks recurring every $1 \, \si{min}$ are observed (Fig.~\ref{fig:ups_driver}b). 
These currents are due to self-tests of the BESS, as confirmed by the manufacturer of this UPS. 
On the AC side, after subtracting the demand of the building connected downstream of the UPS, these $1 \, \si{min}$ peaks are not clearly visible in the three-phase current time series. However, Fourier analysis reveals a persistent $1 \, \si{min}$ component in the currents (Fig.\ref{fig:ups_driver}c). The corresponding periodic power demand can be traced across several transformers and a bus bar, up to the transformer connecting to the $110 \, \si{kV}$ grid (Fig.~\ref{fig:ups_driver}d). These findings demonstrate that the pattern originating on the DC side propagates into the $110 \, \si{kV}$ distribution system.

\begin{figure*}
    \centering
    \includegraphics[width=\textwidth, trim=0cm 0cm 0cm 0cm, clip]{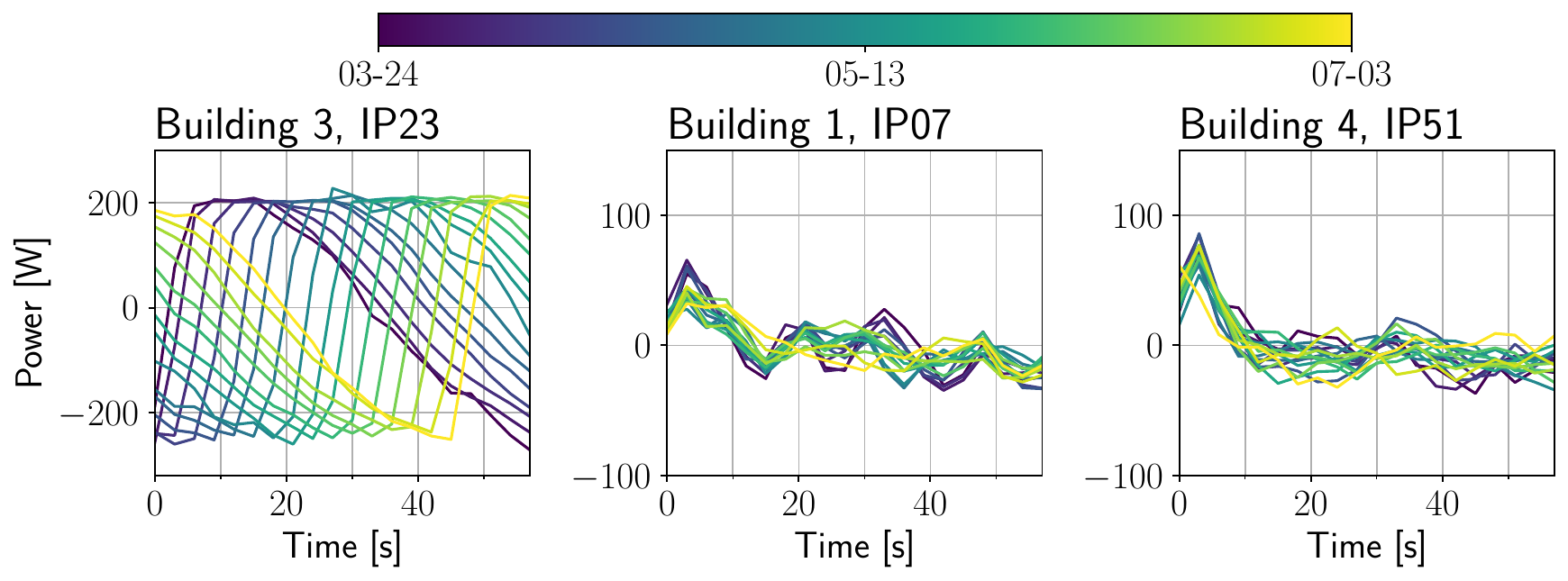}
    \caption{Comparison of weekly wave forms $s_{T = 1 \, \mathrm{week}}(h)$ for three different buildings with known UPS (cf.~Table~\ref{tab:pq_meter_data}) from 2024-03-24 to 2024-07-03. For Buildings 1 (cf. Fig. \ref{fig:fourier_power_campus_16_16}) and 4, the wave forms show a peak in power demand shortly after the beginning of each minute that is stable over the entire measurement period. For Building 3 (cf.~Fig.~\ref{fig:ups_driver}), the wave form is slowly drifting.}
    \label{fig:comp}
\end{figure*}

Finally, we investigate the stability of the pattern and the role of clock synchronization for the operation of the UPS. We first compare the weekly wave forms $s_{T = 1 \, \mathrm{week}}(h)$ of three different transformers (Trafos) serving different buildings that host a UPS system (Fig.~\ref{fig:comp}) and find remarkable differences. 

In Building 3, the wave form drifts slowly over time. In particular, the pattern exhibits a consistent, periodic, sawtooth-like pattern in the active power demand characterized by a sharp increase followed by a gradual descent. This behaviour can be attributed to the system's response to the observed sharp peak in the DC power demand resembling a low-pass filter. 

In contrast, we observe stable wave forms for buildings 1 and 4. These patterns display a peak in power consumption shortly after the beginning of each minute without any drift. Such periodic demand peaks explain the emergence of the $1 \, \si{min}$ pattern in the frequency if we take into account the response of the grid. Remarkably, the UPSs in buildings 1 and 4 have been provided by different manufacturers than the UPS in building 3. This finding emphasizes the crucial role of the design of the battery EMS.

Synchronization of computers connected to the internet is commonly implemented via the Network Time Protocol (NTP)~\cite{mills2002internet}. NTP synchronizes a local clock to the Coordinated Universal Time (UTC) provided by distinguished servers. To directly assess the influence of clock synchronization on the UPS operation in Building 3, we conducted comparative measurements under two conditions: (i) with NTP disabled and (ii) with NTP enabled on the battery EMS.

\begin{figure}
    \centering
    \includegraphics[width=\columnwidth]{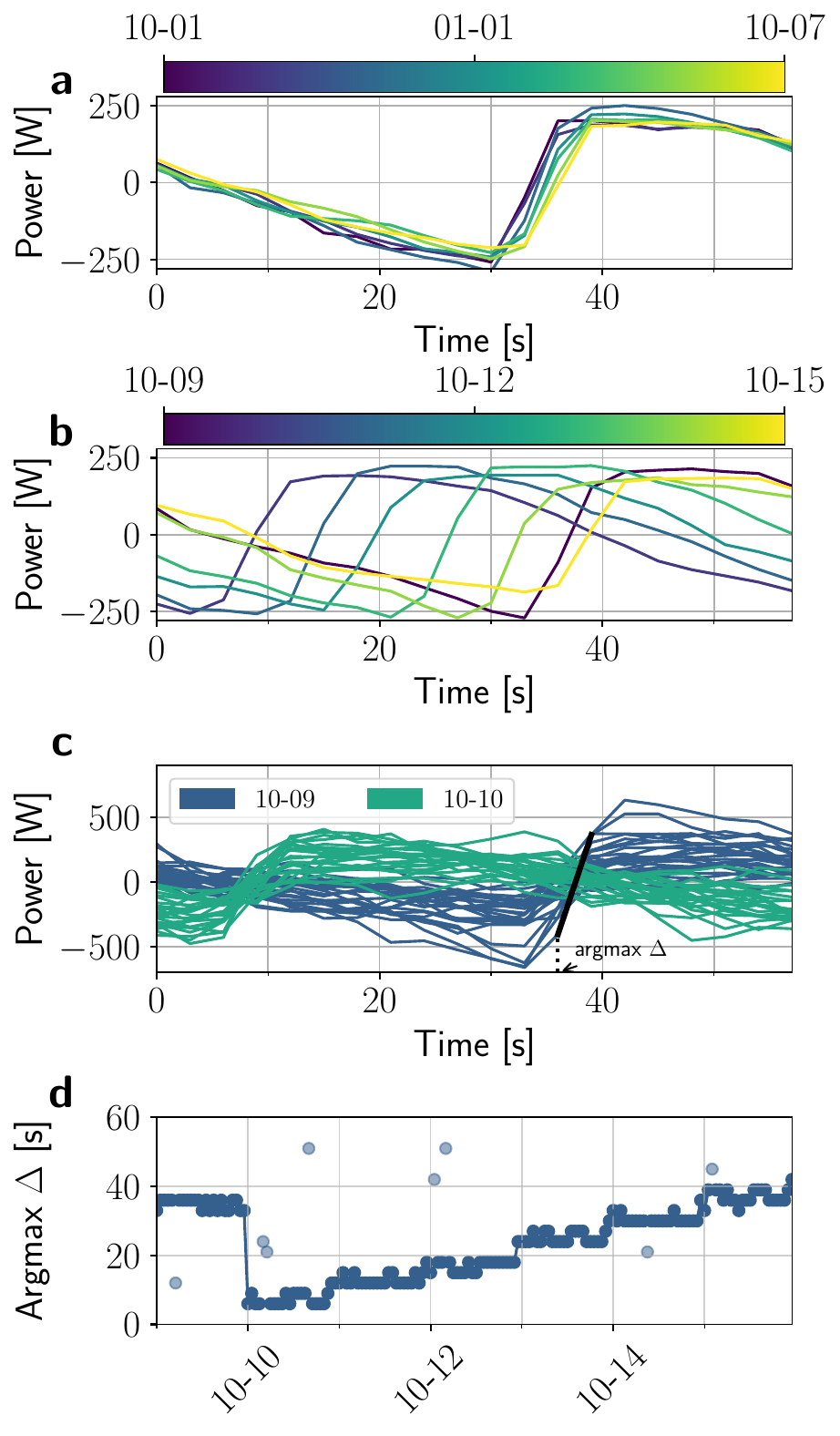}
    \caption{Comparison of the $1 \, \si{min}$ pattern in the active power demand at the transformer level of building 3 hosting a UPS. We compare measurements where the NTP service of the battery EMS is disabled (before 2024-10-08) and enabled (after 2024-10-08).
    \textbf{a} Prior to NTP activation, the daily wave form $s_{T = 1 \, \mathrm{day}}(t)$ exhibited a slow drift. The characteristic sawtooth shape arises from filtering of the discrete current peaks on the DC side of the battery during AC conversion.
    \textbf{b} After NTP was enabled, the drift became significantly more pronounced.
    \textbf{c} Hourly wave forms $s_{T = 1 \, \mathrm{hour}}(t)$ for two representative days following NTP activation show that the phase shift occurs around midnight.
    \textbf{d} Plotting the timing of the sawtooth step (identified as the argmax $\Delta$, see \textbf{c} for a graphical explanation) as a function of time reveals the daily progression after NTP activation.
    }
    \label{fig:rixx_waveform_ntp}
\end{figure}

We find a remarkable difference between the operation with and without NTP. In both cases, the daily wave form $s_{T = 1 \, \mathrm{day}}(t)$ measured at ``Trafo 1'' drifts from day to day -- but the magnitude of the drift differs considerably (Fig.~\ref{fig:rixx_waveform_ntp}a,b). Remarkably, the drift is much larger with NTP enabled, with a magnitude of approximately 5-7 seconds per day. The hourly wave form $s_{T = 1 \, \mathrm{hour}}(t)$ of the power time series at transformer level provides further insights (Fig.~\ref{fig:rixx_waveform_ntp}c,d). We find that the pattern does not drift continuously, but jumps from day to day. An obvious explanation is that the EMS reschedules its operation every day. In summary, self-tests are done periodically in $1 \, \si{min}$ intervals, but the starting point is reset from day to day.

\section{Discussion}

Frequency stability and cybersecurity have been identified as major challenges for the operation of future power systems~\cite{entsoe_frequency,iea_cyber}. Frequency stability is increasingly at risk due to the ongoing decrease in system inertia~\cite{Milano2018FoundationsAC}, while cybersecurity is most commonly discussed in the context of unforeseen errors or malicious attacks~\cite{pasqualetti2013attack}. In this article, we highlight a third emerging challenge that arises from the cyber–physical coupling of modern power systems.

We show that the frequency in various synchronous grids exhibits a recurrence with a period of exactly $1 \, \si{min}$.
While some aspects of this pattern have been recognized before~\cite{oftebro2014analysis,wei2025freqpats}, we provide a comprehensive analysis and identify the key drivers of this pattern. We provide evidence that periodic self-tests of battery energy storage systems in UPSs lead to recurrent peaks in the power demand with a period of $1 \, \si{min}$, leading to a decrease in the frequency. We note that another recurrent disturbance with a period of approx.~$0.07 \, \si{s}$ due to UPSs was recently discovered in Ref.~\cite{mishra2025understanding}.

Our findings clearly show the importance of batteries, but they do not rule out other contributors to the observed $1 \, \si{min}$ pattern. In fact, we observe that the power demand of a smart baking oven also shows periodic peaks (Fig.~\ref{fig:smart_devices}). The oven synchronizes its internal clock via the NTP just as the battery energy management system, leading to a highly stable pattern. Computers are also synchronized via NTP and perform recurring tasks. However, we did not find a $1 \, \si{min}$ pattern in the power load of buildings on the Jülich campus that host a computing centre but do not have a UPS, nor in an ordinary laptop. Furthermore, it appears plausible that water power turbines amplify the pattern due to the excitation of an internal resonance phenomenon~\cite{odelbrink2014design,fcp_project_summary_2017, Pico2012HydroColumbia}.

The periodic variation of the power imbalance is expected to increase as large battery energy storage systems are integrated into the grid 
and smart household devices with synchronized clocks are deployed. In addition, the decreasing system inertia magnifies the frequency variations~\cite{Milano2018FoundationsAC}. We conclude that the amplitude of the $1 \, \si{min}$ pattern is likely to increase further. As the pattern adds up to other deterministic and ambient frequency deviations, it may reach security-relevant levels in future low-inertia grids. 

Fortunately, this trend can be easily mitigated by revised control algorithms that offset the regular self-test of battery energy storage systems. We demonstrated that one manufacturer implemented a daily changing offset, while others maintained a perfectly periodic schedule  (Fig.~\ref{fig:comp}). Randomizing the offset could be even more versatile. In summary, our results highlight the critical importance of robust and secure control methods for future power systems.

\section*{Methods}

\subsection*{Data Sources}

Frequency data on transmission grid level for the CE, GB, and Nordic grids is provided by the transmission system operators. The data has been downloaded and pre-processed using the scripts provided in \cite{pre-processed_scripts}. We use data from the measurement periods 2012-02-01 until 2020-12-31 for the Continental Europe grid, 2015 until 2020 for the Nordic grid and 2014 until 2021 for the GB grid.
Frequency data for Australia, Iceland, South Korea, Malaysia and the USA are measured on the distribution grid level the Electrical Data Recorder described in Ref.~~\cite{jumar2020power,Rydin_Gorj_o_2020fluctuations}. From its point-on-wave data, frequency measurements are derived using a filtered Zero Crossing Method~\cite{forstner2022experimental}. These standalone measurements vary in length and cover approximately 10 months (US West), 2.5 weeks (US East), 3.5 weeks (Australia), 2.5 months (Iceland), 3.5 months (South Korea) and 1.5 weeks (Malaysia). The exact measurement periods are provided in the Figs.~\ref{fig:missing_freq_data}, \ref{fig:missing_freq_data_other_grids_1}, \ref{fig:missing_freq_data_other_grids_2}.

If not explicitly stated otherwise, we use UTC to index the timeseries in order to avoid ambiguity due to daylight saving time. Since we analyse the hourly amplitudes, we further remove full hours of time series data if a single data point is missing or corrupted. The missing data after pre-processing is discussed in appendix \ref{sec:data-missing}.

Inertia data for the GB grid is provided by the National Energy System Operator (NESO)~\cite{NationalGrid_Intertia_data}. The data describes the rotational energy $E_{rot}$ and is given in $\si{GVAs} = \si{GJ}$. It includes the estimated inertia of generators ``with a power output > 15 MW or operating to provide inertia'' and an estimate for demand side contributions~\cite{NationalGrid_Intertia_data}.  Generator data is corrected for actions taken by the National Grid Electricity System Operator including (i) trades to manage constraints or mitigate risks and (ii) instructions to participants in the balancing mechanisms~\cite{NationalGrid_Intertia_data}.

To monitor grid behaviour, load profiles and power flows on the Jülich research campus, over 80 PQI-DA smart power quality (PQ) meters by A-Eberle have been deployed~\cite{pqi-smart}. These Class A devices comply with IEC 61000-4-30 Ed. 3 (2015) standards~\cite{standard_iec_61000_4_30_2015}, ensuring highly accurate measurements of voltage and current at the fundamental frequency.  
Communication with the meters relies on Modbus Transmission Control Protocol (TCP), which provides data via standardised registers for various aggregation intervals, from 200 ms up to 2 hours, as defined in the IEC standards~\cite{standard_iec_61158_2019, standard_iec_61158_6_15_2019}. During normal operation, key electrical quantities such as frequency, voltage, current and power are collected every second and stored on the FIWARE-based Campus Information and Communications Technology (ICT) platform~\cite{Zimmer_2024}. Additional details are provided in the appendix.

\subsection*{Waveform and amplitude}

Let $f(t)$ be the grid frequency data for some discrete time interval $t \in [t_0, t_1]$ with sampling time $\Delta t$. We compute the average wave form  $s_T(t)$ of a recurrent temporal pattern with period $T$ in the interval $[t_0, t_1]$ by averaging over chunks of length $T$. That is, we divide the interval $[t_0, t_1]$ into $ N = \frac{t_1 - t_0}{T_{pattern}}$ blocks of length $T_{pattern}$ and average $f(t)$ over these blocks
\begin{equation}
    s_T(t, t_0, t_f) := \frac{1}{N} \sum_{n=0}^{N-1} f(n T + t) \quad \text{ for } t \in [0,T).
    \label{eq:signalform}
\end{equation}
Similarly, the wave form can be computed using harmonic analysis. Starting with the Discrete Time Fourier (DFT) decomposition $\mathcal{F}(f,T)$, we keep only the components with the desired periodicity and the corresponding higher harmonics while discarding other Fourier components.
We then apply the inverse DFT to obtain the filtered pattern in the time domain. Both methods give identical results. From the wave form, we define the amplitude $a_T$ of the pattern in the interval $[t_0, t_1]$ as
\begin{equation}
    \label{eq:Tennet_Amplitudes}
    a_{T}\big([t_0, t_f] \big):= \frac{1}{2} \bigg( \max_{t \in [0,T)} s_T(t, t_0, t_f) - \min_{t \in [0,T)} s_T(t, t_0, t_f) \bigg).
\end{equation}

To quantify how wave forms change over time, we divided the entire measurement period into one-hour segments and evaluated the wave form and the amplitudes separately for each segment. That is, for each hour $h$ we set $t_0 = h$ and $t_1 = h + 1 \, \si{hour}$. The resulting hourly wave form is denoted as $s_h(t)$ and the hourly amplitudes are denoted by $a_h$. 

Based on the hourly amplitudes, we can also quantify characteristic variations in the strength of the $1 \, \si{min}$ pattern. In particular, we evaluate the daily profile by averaging over all days during the measurement period. Let $h$ denote the $h$-th hour in a week, the daily profile is defined as 
\begin{equation}
   a_{T,\mathrm{daily \, profile}}(h) = \frac{1}{N_{days}}\sum_{d=0}^{N_{days}-1} a_{T}(h + 24 d),
   \label{eq:amplitudes_weekly_profile}
\end{equation}
where $N_{days} = \lfloor \frac{(t_f - t_0) \Delta t}{3600 \cdot 24} \rfloor $ is the number of days in the time interval $[t_0, t_f]$ and $\lfloor \cdot \rfloor$ denotes rounding towards minus infinity. The weekly profile is calculated similarly. 

\subsection*{The aggregated swing equation}

The temporal evolution of the grid frequency is described by the aggregated swing equation (ASE)~\cite{Ulbig2013Inertia_swing_equation,Bialek2020stabilityandcontrol},
\begin{align}
   \frac{2 E_{\textrm{rot}}}{f_{ref}} \frac{d f}{d t} =  \Delta P(t) - D f(t).
    \label{eq:swing_equation_general}
\end{align}
Here, $\Delta P$ is the power imbalance, $D$ is a damping constant, and $f_{ref}$ is the reference frequency. The inertia is quantified by the rotational energy $E_{\textrm{rot}}$ of all online synchronous machines. Neglecting the damping, the Rate of Change of Frequency $df/dt$ is proportional to the power imbalance $\Delta P$ divided by the inertia.

If we have data for the hourly inertia $E_{rot}(h)$ in the grid, we can approximate the hourly power amplitudes  $P_{T}(h)$ for any pattern. To this end, we insert the following ansatz
\begin{align*}
    f(t,h) &= a_T(h) \sin \left(
    \frac{2 \pi t}{T} \right), \\
    \Delta P(t,h) &= P_{T}(h) 
    \cos \left( \frac{2 \pi t}{T} \right). 
\end{align*}
into the ASE. For simplicity, we linearize the equation and neglect the damping (i.e.~$D=0$) and thus get 
\begin{equation}
    P_{T}(h) =  a_T(h) \cdot \frac{4 \pi E_{rot}(h)}{f_{ref} T}.
    \label{eq:power_amplitudes}
\end{equation}
We note that this treatment captures the inertial response of the grid, but not the presence of potential resonance phenomena, including the resonance of hydro power turbine systems.

\section*{Acknowledgements}

We gratefully acknowledge the funding we received from the Helmholtz Association and the Federal Ministry for Research, Technology and Aeronautics (BMFTR) for project 03SF0628, which funded the measurement infrastructure. B.S. and U.O. acknowledge funding from the Helmholtz Association under grant no. VH-NG-1727 and from the Helmholtz Networking Fund through Helmholtz AI.

\clearpage

\appendix

\section{The $1 \, \si{min}$ pattern in Mallorca}
\label{sec:Mallorca}

\begin{figure}[!h]
    \centering
    \includegraphics[width=\columnwidth]{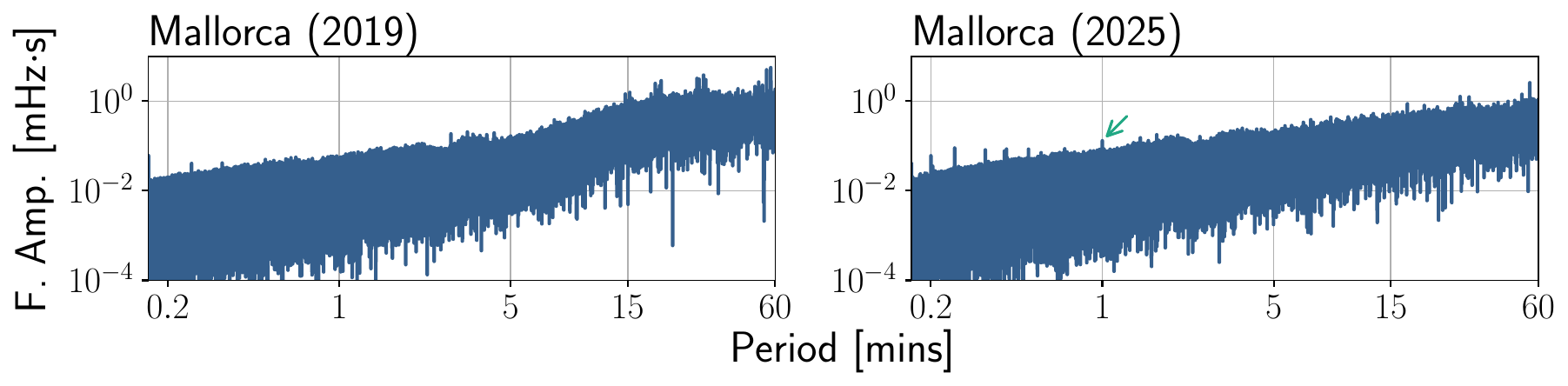}
    \caption{Fourier spectrum of the power grid frequency for Mallorca, an island grid in Spain. The measurement periods are given in  Fig.~\ref{fig:missing_freq_data_other_grids_2}.}
    \label{fig:fourier_no_peaks_freq}
\end{figure}

Not all synchronous grids exhibit, or have exhibited, a $1 \, \si{min}$ pattern in the grid frequency in recent years. We consider the grid frequency on the Spanish island Mallorca (Fig.~\ref{fig:fourier_no_peaks_freq}). The Majorcan grid is connected to the Continental European (CE) synchronous area via a high-voltage direct current (HVDC) link only. This allows for power transmission but does not establish frequency synchronization between the two systems.

Interestingly, while no evidence of a $1 \, \si{min}$ pattern was found in Mallorca in late 2019, our measurements in 2025 clearly reveal the presence of this pattern. 
This finding suggests that potential drivers---most notably large-scale battery energy storage systems (BESS)---were not yet present on the island in 2019. The appearance of the $1 \, \si{min}$ pattern in 2025 aligns with reports of BESS installations in the Balearic Islands in recent years~\cite{nhoa_enerst_suppli_105mwh_2024, red_electrica_grid_booster_2025}, supporting the hypothesis that such systems are key contributors to this characteristic pattern.

\section{Wave forms in major grids}

\begin{figure}[bt]
    \centering
    \includegraphics[width=1\columnwidth]{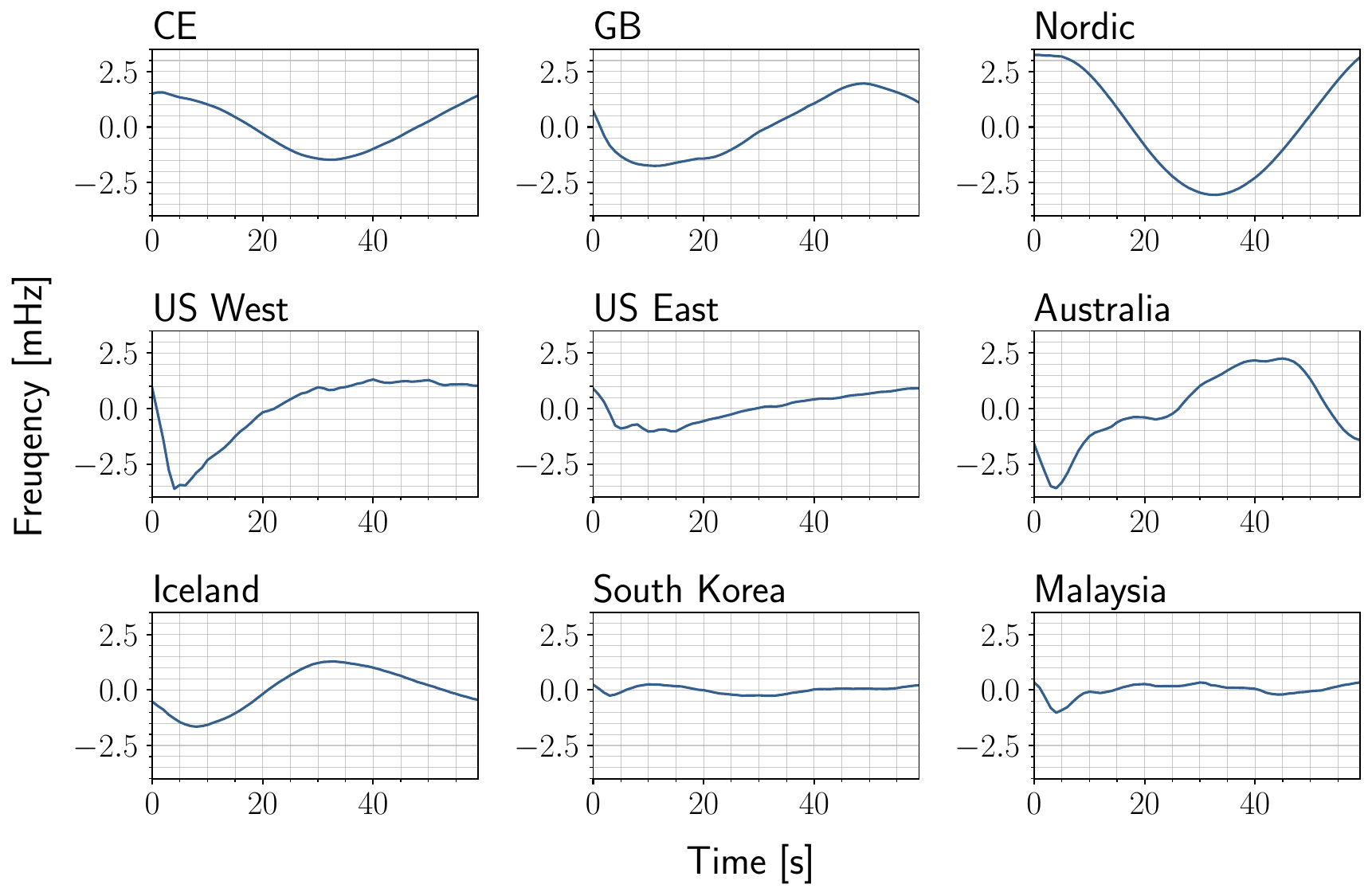}
    \caption{Comparison of wave forms for all grids that exhibit a $1 \, \si{min}$ pattern averaged over all available data. }
    \label{fig:waveforms_freq_grids}
\end{figure}

Average wave forms for all grids that exhibit a $1 \, \si{min}$ pattern are shown in~Fig.~\ref{fig:waveforms_freq_grids}. For most grids (GB, US East and West, Australia, Iceland, South Korea, and Malaysia), the frequency drop occurs at the start of each full minute. This initial drop supports the hypothesis that the pattern is driven by recurring peaks of the power demand in the beginning of the minute associated with clocked control algorithms. In the Continental European (CE) and Nordic grids, the frequency also decreases shortly after the start of each hour, but more gradually, reaching a minimum after approximately $30 \si{s}$.

\section{The impact of Covid-19 on the $1 \, \si{min}$ pattern in the \textit{CE} grid}

\begin{figure}[tb]
    \centering
    \includegraphics[width=1\columnwidth]{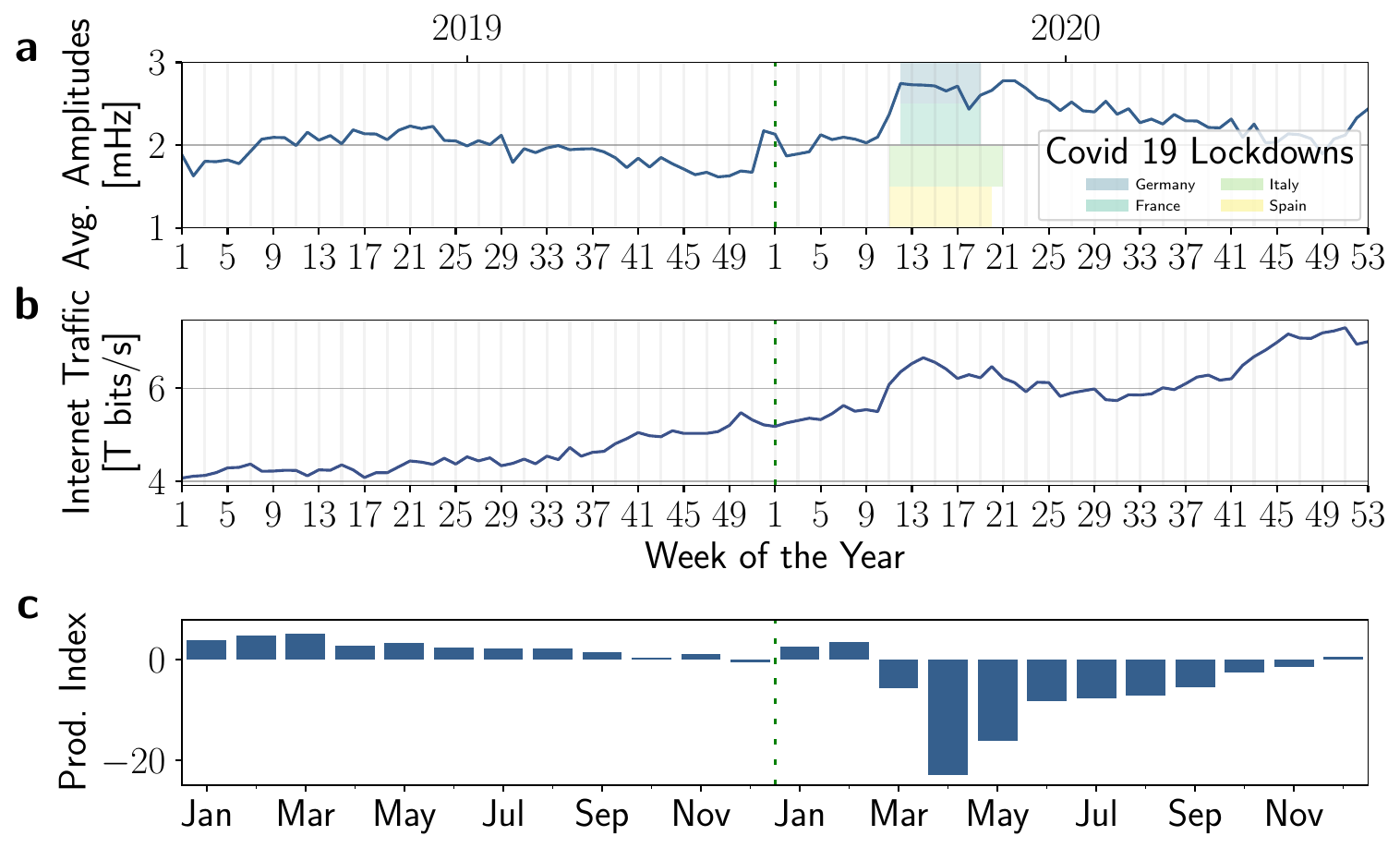}
    \caption{
    Impact of the Covid-19 lockdown on the $1 \, \si{min}$ pattern in comparison to internet traffic and industry production.
    \textbf{a}: The weekly average hourly amplitudes of the $1 \, \si{min}$ pattern increases substantially at the beginning of the lockdown.
    \textbf{b}: The figure shows the internet traffic at the DE-CIX node in Frankfurt, Germany, which is the largest node in Europe \cite{de-cix-statistics}. The traffic increased significantly at the beginning of the lockdowns.
    \textbf{c}: The industry production decreased significantly during the lockdown. The figure shows the production index of the German industry in reference to 2015, which is monthly evaluated by the German statistical office \cite{destatis-production}.
    }
    \label{fig:amplitude-lockdown}
\end{figure}

The Covid-19 pandemic had a strong impact on the economic activities and daily life across the globe and thus also affected the way electric power is consumed. The pandemic reached Europe in early 2020 and many countries issued a lockdown to suppress new infections in spring 2020. Figure~\ref{fig:amplitude-lockdown} illustrates two characteristic aspects of the lockdown in Germany in April 2020. Industry production decreased substantially, in particular in the automotive sector. At the same time the internet traffic increased strongly, for instance, due to extended home office or video streaming. 

The weekly averaged amplitudes of the $1 \, \si{min}$ pattern increased strongly by a factor of $1.3 - 1.4$ at the beginning of the lockdown. There is no corresponding decrease at the end of the lockdown, as this date coincides with the seasonal maximum discussed above. The amplitude remains high and decreases only later during the year. 

We conclude that it is unlikely that industry processes are the main cause of the $1 \, \si{min}$ pattern. During the lockdown, industrial activities decreased strongly, while the amplitude of the $1 \, \si{min}$ increased. In contrast, industry processes may contribute to the damping of frequency variations by providing inertia. 

The utilization of information and internet infrastructures, measured by the internet traffic at the DE-CIX internet node, shows a strong increase during the  lockdown, just as the pattern amplitude. Large computing centres typically operate uninterruptible power supplies (UPSs) with BESSs to avoid data loss in the case of a power outage. Hence, it appears possible that increased internet traffic correlates with a stronger use of UPSs.

\section{Geographical Analysis of the $1 \, \si{min}$ pattern in the \textit{CE} grid}

We now compare frequency recordings from various locations in the \textit{CE} grid for one week in July 2019, which may provide hints towards the causes of the $1 \, \si{min}$ pattern. For example, if the source was localized, we expect the pattern to be strongest in the vicinity of the source. We thus extract the amplitudes and average wave forms of the pattern from synchronized measurements for all locations, as these are our main statistical indicators for the $1 \, \si{min}$ pattern. 

\begin{figure}[tb]
    \centering
    \includegraphics[width=\columnwidth]{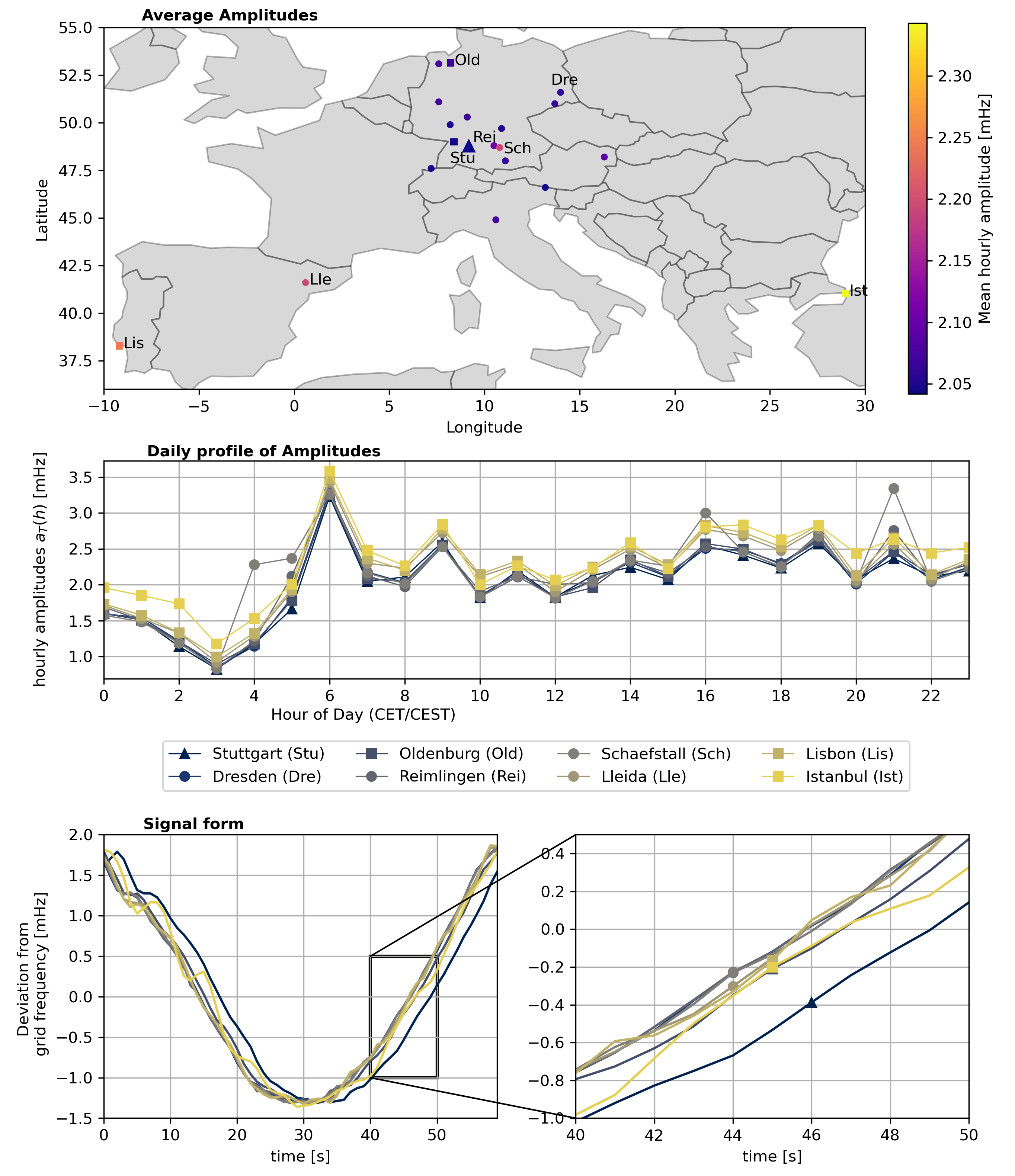}
    \caption{Geographic features of the $1 \, \si{min}$ pattern for one week in June 2019, combining data from KIT, GridRadar and TransNetBW. We join the datasets and only consider those hours for which we have full data available. \textbf{a}: Map of synchronous measurement locations. Circles correspond to recordings from GridRadar \cite{GridRadar}, rectangles correspond to recordings from KIT~\cite{jumar2020power} and the triangle corresponds to the measurement on the transmission grid level by TransNetBW \cite{TransnetBW_data}. The color indicates the   strength of the average amplitudes. \textbf{b}: The daily profiles of the amplitudes $a_{T = 1 \si{min}}(h)$ for several selected measurement locations reveal that the pattern is stronger in Scheafstall for several hours. \textbf{c}: Comparison of the wave forms for several selected locations. The pattern lags behind in the transmission grid.}
    \label{fig:geographic_distribution_2019}
\end{figure}

Most recordings have been performed on ordinary pocket sockets on the distribution grid level, among those are the recordings obtained by scientists of the Karlsruhe Institute of Technology (KIT) in Lisbon (Portugal), Karlsruhe, Oldenburg (Germany) and Istanbul (Turkey)~\cite{jumar2020power} and by the company \textit{GridRadar} in various locations mostly in Germany~\cite{GridRadar}, cf.~map in Fig.~\ref{fig:geographic_distribution_2019}. We compare these frequency recordings with measurements on the transmission grid level performed by the TSO \textit{TransNetBW} in Stuttgart~\cite{TransnetBW_data}. The measurement period covers one week in July 2019, but the KIT dataset is missing some recordings for some hours. We exclude these hours when we compare the locations between different datasets.

\begin{figure}[tb]
    \centering
    \includegraphics[width=\columnwidth]{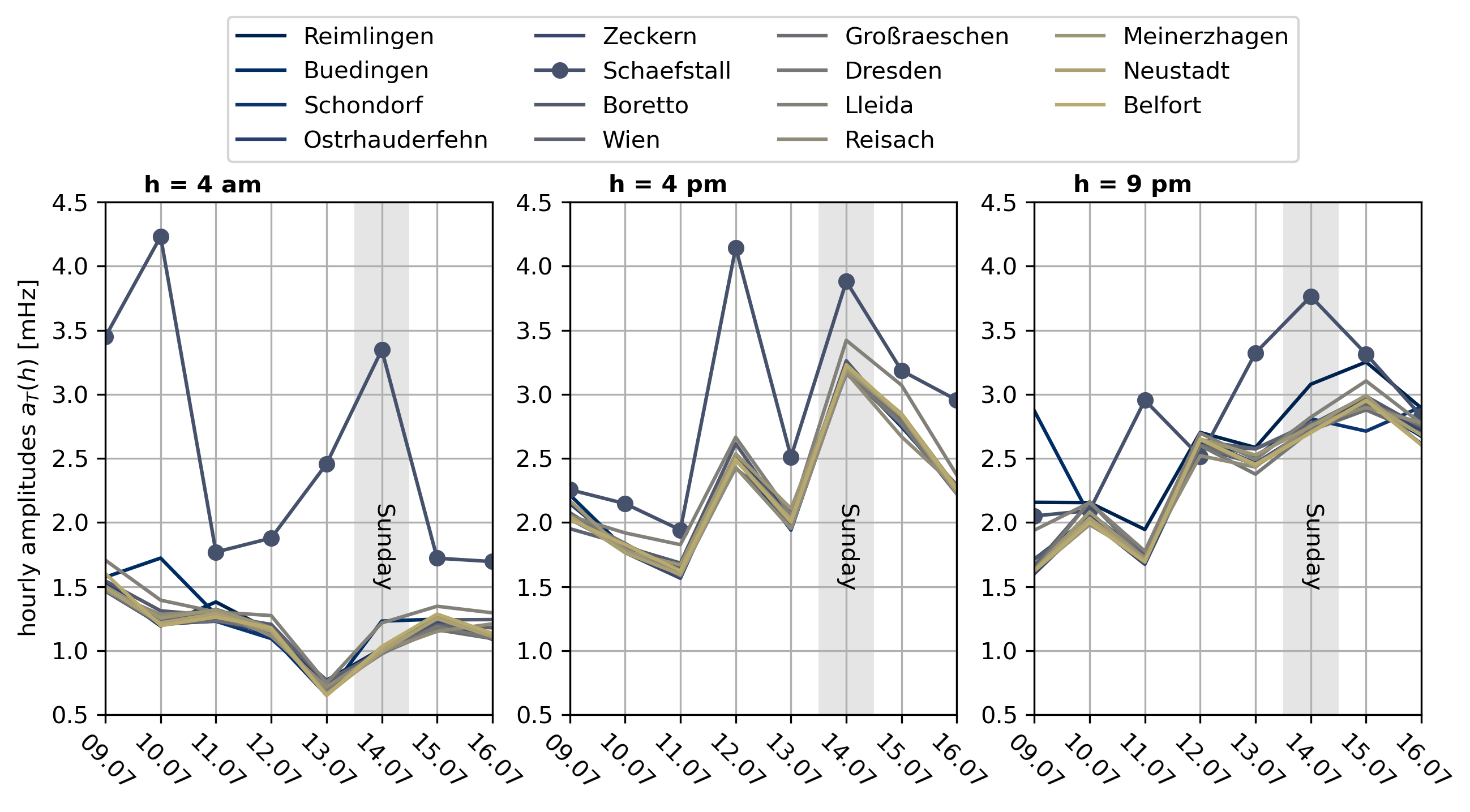}
    \caption{Comparison of the hourly amplitudes at 4 am, 4 pm and 9 pm over the full week between the recordings performed by GridRadar \cite{GridRadar}. For these hours, the hourly amplitudes are (almost) constantly higher in Schaefstall than in the other locations. Thus, a measurement error appears unlikely.}
    \label{fig:geographic_distribution_2019_grid_radar}
\end{figure}

The $1 \, \si{min}$ pattern shows some interesting regional differences. The mean hourly amplitudes $\langle a_{T = 1 \si{min}}(h) \rangle$ are generally stronger in the periphery of the grid, that is in Istanbul and Lisbon, compared to the core in Germany, cf.~Fig.~\ref{fig:geographic_distribution_2019}a. This is consistent with  earlier studies \cite{Rydin_Gorj_o_2020fluctuations} on spatial features of frequency fluctuations. Fluctuations in the outskirts of the \textit{CE} grid are generally stronger than in the centre. However, measurements in Schaefstall in Bavaria, Germany, break this rule. The average amplitudes are significantly stronger compared to the other locations in the vicinity, cf.~Fig.~\ref{fig:geographic_distribution_2019}a.  
To further investigate this `Schaefstall Anomaly' we plot the average amplitudes over the course of a day, cf.~Fig.~\ref{fig:geographic_distribution_2019}b. The daily profiles reveal that the anomalous high average amplitude in Schaefstall stems from high amplitudes at certain hours of the day, in particular $4$ am, $4$ pm and $9$ pm. Furthermore, the amplitudes in Schaefstall are (almost) constantly higher than at the other locations for these hours, cf.~Fig.~\ref{fig:geographic_distribution_2019_grid_radar}. This suggests that the anomaly is not caused by a faulty measurement by the recorder in Schaefstall and thus might give a hint to potential local drivers or amplifying factors due to a resonant behaviour.

\begin{figure}[tb]
    \centering
    \includegraphics[width=1\columnwidth]{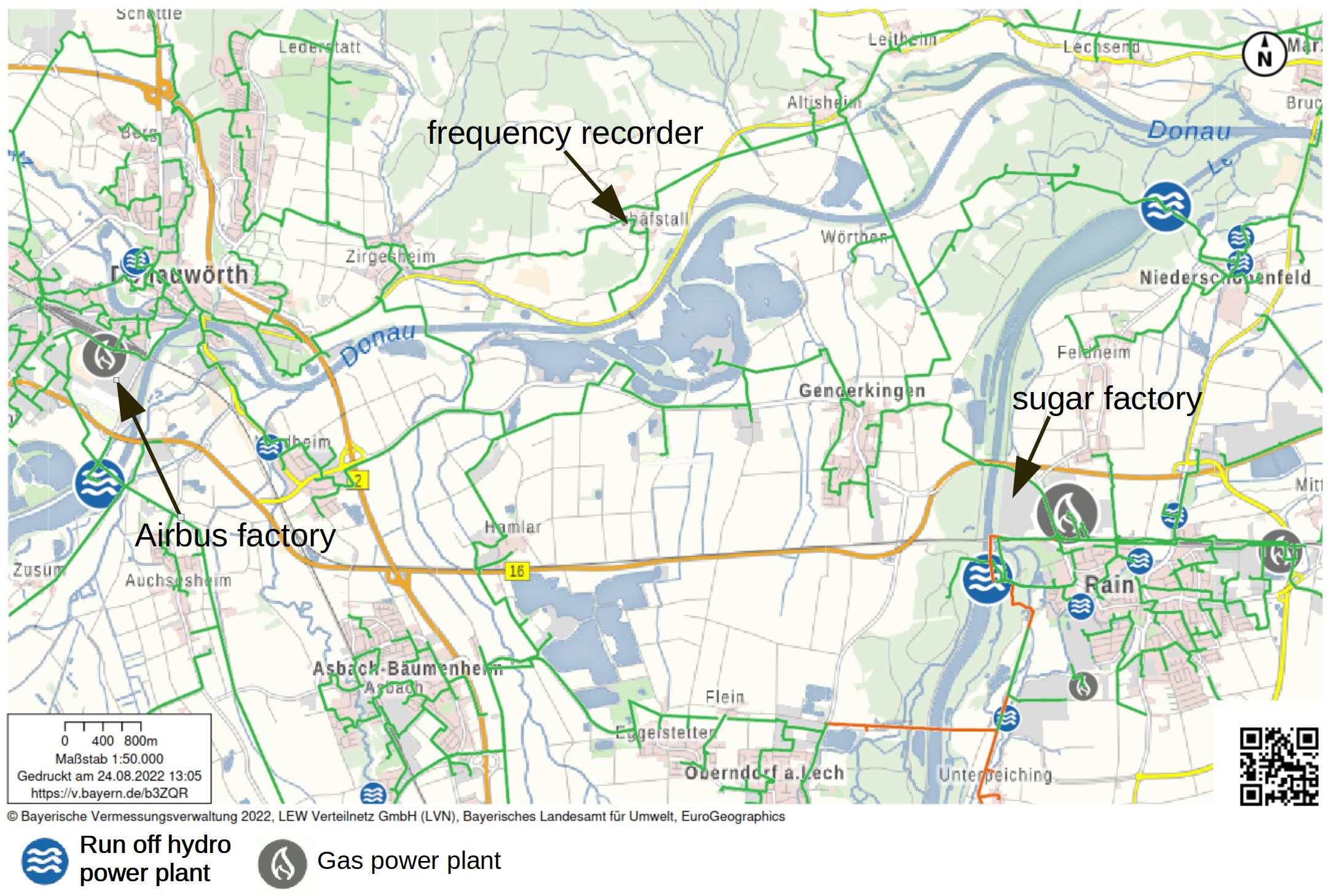}
    \caption{Map of points of Interest around Schaefstall, reproduced from Energie Atlas Bayern \cite{Energie_Atlas_bayern} with permission. There are several factories, gas power plants and run-of-river hydro power plants that could be potential local drivers or amplifiers of the $1 \, \si{min}$ pattern. }
    \label{fig:map_Schaefstall}
\end{figure}

To identify potential local contributing factors, we investigate which generators and consumers are in the vicinity of the frequency recorder in Schaefstall, see Fig.~\ref{fig:map_Schaefstall}. There are two large factories, a sugar factory in Rain and an Airbus Helicopters factory in Donauwoerth, that are connected to the local distribution grid according to~\cite{Energie_Atlas_bayern}. Each has its own gas power plant.  Furthermore, we find several run-of-river hydro plants on the side rivers of the Danube. These, among others, could be potential local driving or amplifying factors.

Hydro power plants have been identified as contributing factors of slow frequency oscillations~\cite{oftebro2014analysis,odelbrink2014design,fcp_project_summary_2017, Pico2012HydroColumbia}. These studies suggest that slow periodic oscillations are related to the resonance effects in the primary frequency control performance of the hydraulic turbine control system. Hence, the resonant behaviour of hydro turbines can explain a local amplification of the amplitude of the $1 \, \si{min}$ pattern. 

Finally, we also consider the respective local wave forms, which show slight differences in the phase shift, increasing from West to East (Fig.~\ref{fig:geographic_distribution_2019}). However, all measurements have been performed on ordinary pocket sockets in the low-voltage distribution grid, impeding the interpretation of the results. Indeed, a much stronger difference is observed between recordings from the distribution grid in Karlsruhe and from the TSO TransNet BW in close geographic proximity. Remarkably, the pattern extracted from the TransNet data is lagging behind. This may hint at a pattern that is rooted in the distribution grid, with the transmission grid being driven and following with a certain lag.

\section{Power Measurements in Juelich}

This section presents the analysis conducted on the data obtained from the FZJ campus electrical grid. After a brief overview of the electrical system and the measurement infrastructure, the Fourier spectra of active and reactive power for selected significant locations are provided.

\subsection{Forschungszentrum Jülich Electrical Grid}

The electrical grid of the Jülich research campus consists of multiple medium voltage (MV) distribution feeders at 10 kV, the majority of which are reclosing on the same busbar at the primary substation. This configuration forms ring topologies, which are typically employed to increase the resiliency of the grid. 
At the different nodes of the feeders, the medium voltage (MV, 10 kV) to low voltage (LV, 0.4 kV) transformers connect to office and laboratory buildings, which could also be equipped with a UPS including a BESS or photovoltaic power sources. Moreover, the electrical grid of the campus is characterized by large-scale distributed energy resources, including a large photovoltaic power plant. Fig.~\ref{fig:meas_system} provides a schematic representation of a portion of one of the campus's distribution rings where buildings are connected to the MV level via transformers.

\subsection{Measurement Infrastructure of the Campus}

The electrical grid of the FZJ campus is continuously monitored at multiple locations, with over 80 PQI-DA smart power quality (PQ) meters manufactured by the company a-eberle~\cite{pqi-smart}. These devices, which are in compliance with IEC 61000-4-30 Ed.3 (2015) standard for an A-Class device~\cite{standard_iec_61000_4_30_2015}, are capable of providing synchronized measurements of several electrical quantities with an accuracy of $0.1 \%$ at the fundamental frequency. By means of the Modbus TCP protocol, different registers of the PQ meters are used to collect voltage, current, phase angle and frequency from different locations of the campus.
The timestamped and synchronized measurements are acquired with 3 second time interval and stored on the campus FIWARE-based ICT platform~\cite{Zimmer_2024}, which allows for a smooth extrapolation of the data by selecting the time period and data resolution. 

\begin{figure}[t]
    \centering
    \includegraphics[width=0.9\columnwidth]{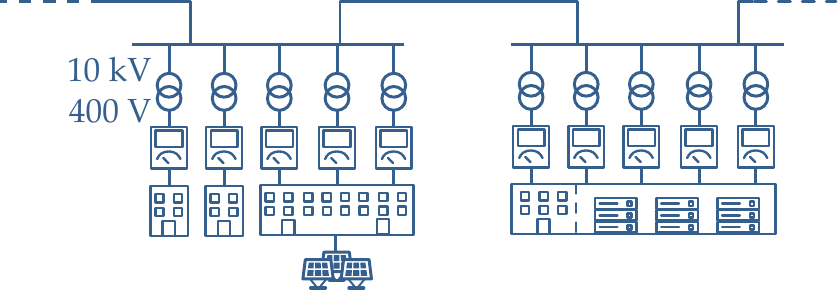}
    \caption{Schematic representation of the Measurement Infrastructure of the Campus}
    \label{fig:meas_system}
\end{figure}

As previously explained, the campus electrical grid is composed of several main MV distribution feeders, to which one of multiple buildings are connected by means of MV/LV transformers. Some large buildings use multiple transformers to supply different portions of the building due to high load demands. An explanatory description of the 10 kV MV distribution grid of the campus and the connection to the 400 V LV system is presented in Fig.~\ref{fig:meas_system}, which shows several MV/LV transformers connected to the same MV bus to supply one or more buildings.
For this reason, the data from multiple PQ meters has been used to characterize the power quality and calculate the Fourier spectra for some of the campus' buildings. 

For this analysis, the data of two different periods have been considered:
\begin{itemize}
    \item MP1: from 2024-03-24 00:00:00 to 2024-07-03 23:59:57
    \item MP2: from 2024-10-01 00:00:00 to 2024-11-05 23:59:54
\end{itemize}

In addition, the data collected from a high-power BESS (1.5 MW / 0.5 MWh) installed in the campus has been considered. The BESS is permanently connected via a dedicated transformer to the electrical grid, and it is typically operated as a UPS. As described in the main manuscript, the UPS topology exhibits three battery racks connected in parallel to the DC link between the grid-side converter and the building-side converter.

\subsection{Result of the Data Analysis of the Campus}

\begin{figure}[tb]
    \centering
    \includegraphics[width=\columnwidth]{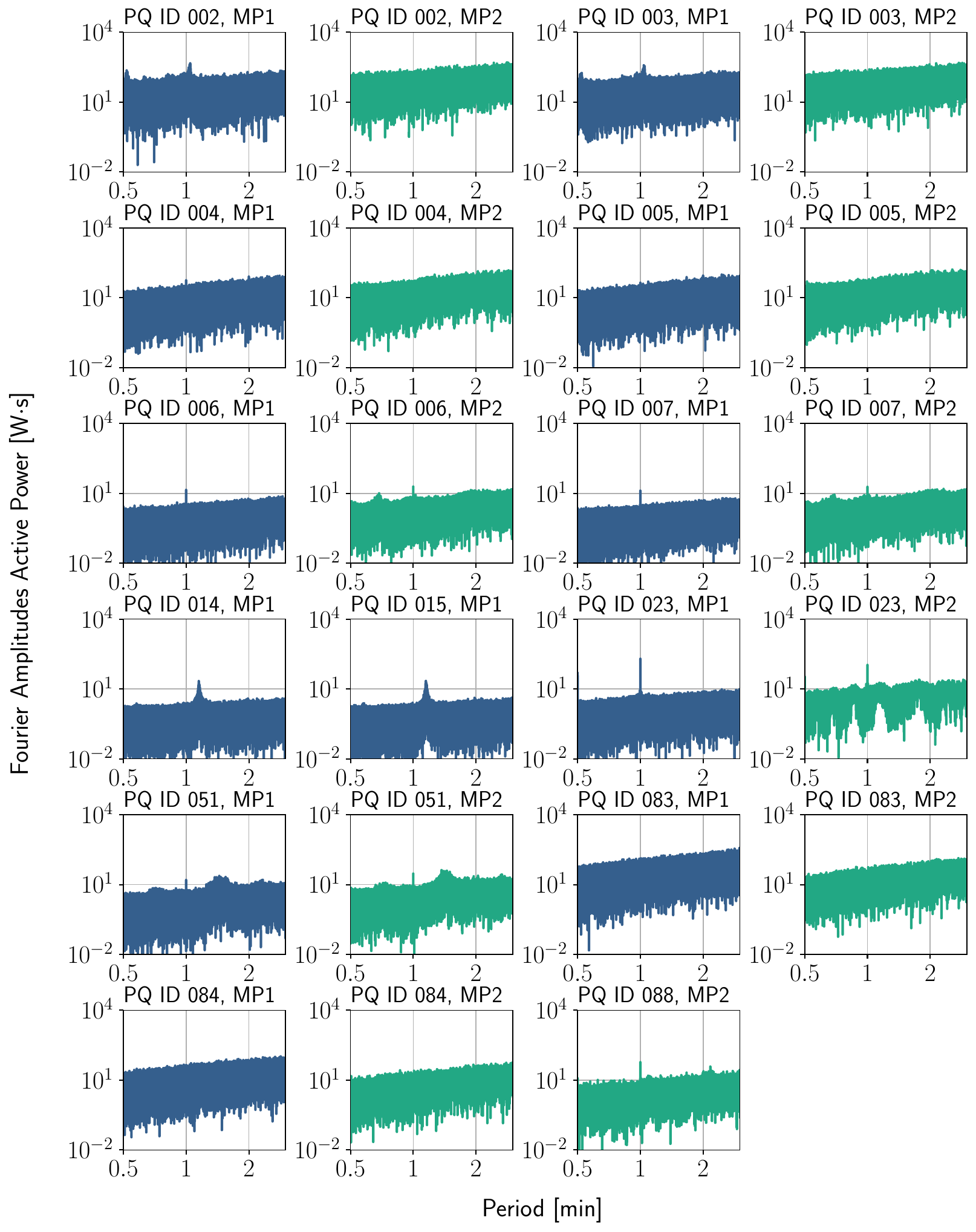}
    \caption{Fourier Spectra of Active Power $P$ for selected PQ Meters on FZJ campus for different measuring periods. Blue is MP1 and green is MP2. }
    \label{fig:campus_active_fourier}
\end{figure}

\begin{figure}[tb]
    \centering
    \includegraphics[width=\columnwidth]{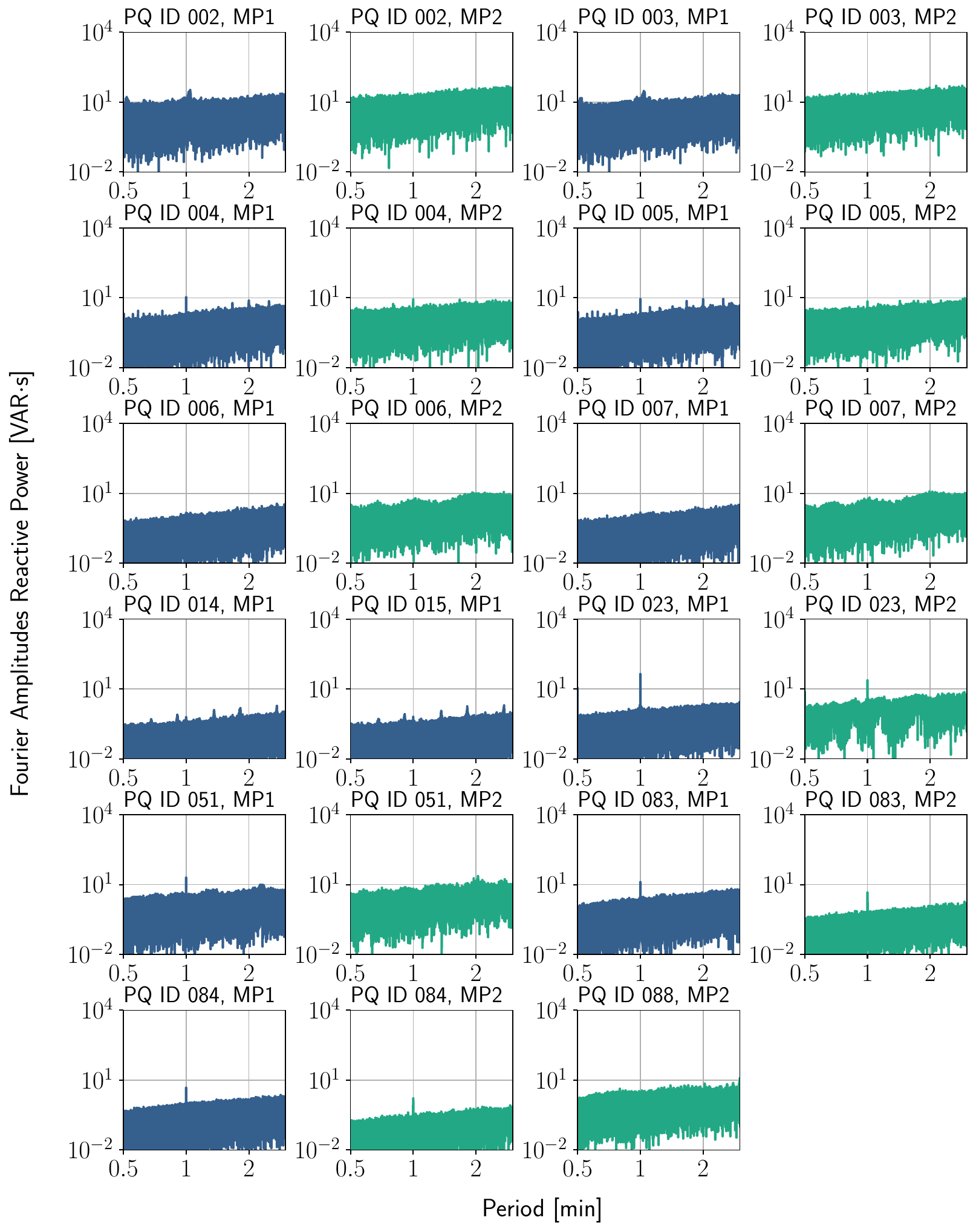}
    \caption{Fourier Spectra of Reactive Power $Q$ for selected PQ Meters on FZJ campus for different measuring periods. Blue is MP1 and green is MP2. }
    \label{fig:campus_reactive_fourier}
\end{figure}

For each building of the campus where PQ meters are installed, the measurements of active and reactive power from all the PQ meters associated with the same building have been collected for both time periods, MP1 and MP2. Then, the Fourier spectra have been calculated for each set of measurements, as displayed for the most significant ones in Fig.~\ref{fig:campus_active_fourier} and Fig.~\ref{fig:campus_reactive_fourier} for active and reactive power, respectively. We evaluate if the spectra show a peak at $1 \, \si{min}$ or other significant patterns. The findings are summarized in Table \ref{tab:pq_meter_data}. 

\begin{table*}[tb]
    \centering
    \begin{tabular}{c|c|c|c|c|c|c}
        PQ ID & Building & \multicolumn{2}{c|}{Data} & \multicolumn{2}{c|}{$1 \, \si{min}$ Pattern} &  Description  \\
        &  & MP1 & MP2 & $P$ & $Q$ &  \\
        \hline
        002 & 1 & \checkmark & \checkmark & - & - & High Performance Computer \\
        003 & 1 & \checkmark & \checkmark & - & - &  High Performance Computer \\
        004 & 1 & \checkmark & \checkmark & MP1 & both &  UPS, Offices \\
        005 & 1 & \checkmark & \checkmark & MP1 & both &  UPS, Offices \\
        006 & 1 & \checkmark & \checkmark & both & - &  UPS, Offices \\
        007 & 1 & \checkmark & \checkmark & both & - &  UPS, Offices \\
        014 & 2 & \checkmark & - & - & MP1 &  Offices \\
        015 & 2 & \checkmark & - & - & MP1 &  Offices \\
        023 & 3 & \checkmark & \checkmark & both & both &  UPS ({1.5} {MW} / {0.5} {MWh}) \\
        051 & 4 & \checkmark & \checkmark & both & MP1 &  UPS, Offices, PV, BEV charging \\
        083 & 5 & \checkmark & \checkmark & - & both & PV field \\
        084 & 5 & \checkmark & \checkmark & - & both & PV field \\
        088 & 6 & - & \checkmark & MP2 & - & UPS, Offices \\
    \end{tabular}
    \caption{Overview of all PQ Meters with interesting patterns in the active $P$ (cf.~Fig.~\ref{fig:campus_active_fourier}) or reactive power $Q$ (cf.~Fig.~\ref{fig:campus_reactive_fourier}). We compare two measurement periods, MP1 and MP2. The data columns indicate for which PQ Meter we collected data in the two periods.
    }
    \label{tab:pq_meter_data}
\end{table*}

The results indicate that the Fourier spectrum for the PQ meter 023 has a very sharp peak at 1 minute for both active and reactive power in both data sets. This assertion is supported by the fact that the measurement is collected after the MV/LV transformer that connects the high-power BESS (1.5 MW / 0.5 MWh) that currently serves as a UPS. This device, as described in the main paper, injects current peaks with a 1 minute period into the grid, thereby generating the distinct peak in the spectrum. 

Similarly, the same type of peaks are visible in measurement sets of other buildings. Specifically, it can be noticed that the PQ meters 006, 007, 051, 088 show a distinct peak at 1 minute for active power. Notably, PQ meter 088 was online only in measurement period MP2. These four PQ meters monitor the transformers supplying different buildings that all host one or multiple UPSs based on BESSs. 

The PQ meters 004 and 005 show small peaks at 1 minute for active power during the measurement period MP1. These peaks are hardly visible on a logarithmic scale as in Fig.~\ref{fig:campus_active_fourier}, but clearly visible on a linear scale. We do not observe comparable peaks during the measurement period MP2, though. These two PQ meters monitor the transformers supplying one building, which hosts one UPS based on a BESS. 

In addition, we observe broad peaks in the Fourier spectrum of the active power consumption at approximately 70 seconds for the PQ meters 014 and 015 supplying typical office buildings. The strong dispersion of the peaks points towards a recurrent pattern that is either short in duration, strongly damped or subject to noise or jitter. In any case, this signature strongly differs from the stable periodic patterns observed for other PQ meters at 1 minute.

Finally, the Fourier spectrum of the reactive power time series shows a peak at 1 minute for several PQ meters. These PQ meters serve different types of buildings as well as photovoltaic power sources. These patterns deserve further  investigation.

\section{Background on Battery Energy Management Systems}

Regulatory requirements and market conventions play a central role in shaping the temporal architecture of battery energy management systems (EMS) and aggregated virtual power plants (VPPs).  
In Europe, Frequency Containment Reserve (FCR) frameworks established by ENTSO-E and national transmission system operators (TSOs) mandate full active power delivery within 30–60 seconds after activation \cite{entsoe_lfcr_annex2023,statnett_fcr_2022}.  
These stringent dynamic response requirements impose a fundamental design trade-off on EMS: systems must react quickly enough to grid events while maintaining internal stability and minimizing degradation.

When setpoints are updated simultaneously across many distributed units, the resulting synchronized power responses can leave measurable, periodic imprints in system frequency \cite{hoke2021,kruse2020}.  
Such deterministic patterns have indeed been observed in frequency deviation analyses conducted by ENTSO-E \cite{entsoe_dfd2020}.

At the low-level control layer, EMS continuously monitor grid frequency and voltage, typically at sampling rates between 10 Hz and 100 Hz to capture short-term dynamics \cite{tang2023}.  
Higher-level dispatch, optimization, and scheduling processes, however, are generally executed in coarser, discrete intervals.  
This cadence reflects the operational design of EMS and distributed energy resource management systems, as documented in multiple deployments and virtual power plant studies \cite{hoke2021,kung2016_supervisory}, and naturally leads to some temporal alignment across assets.

Within individual EMS architectures, internal functions such as state-of-charge (SOC) balancing, ramp-rate adjustment, and recharge management often follow minute-based or multi-minute cycles.  
These scheduled internal updates help suppress overreactions to high-frequency noise while ensuring stable control performance \cite{gundogdu2017,hoke2021}.  
Similarly, some industrial battery and UPS controllers perform automated health checks or brief diagnostic pulses at a one-minute interval to maintain voltage and performance stability \cite{ups_blog,snaptec_akkutec_manual}.  
Across different application domains, this convergence of update intervals may contribute to recurring minute-scale patterns in the frequency spectrum, hinting that synchronized EMS updates can influence grid dynamics.

Empirical frequency measurements frequently reveal small but recurring deviations at roughly one-minute periods, consistent with collective EMS updates or synchronized dispatch cycles \cite{kruse2020,entsoe_dfd2020}.  
These observations underscore the importance of explicitly considering EMS timing and coordination when analyzing system-wide stability and frequency behavior.  
Although minute-scale supervisory synchronization is often convenient for communication and market alignment, it can introduce periodic forcing that interacts with natural system modes, thereby shaping emergent grid dynamics.

\section{Smart Household Devices}

\begin{figure}[tb]
    \centering
    \includegraphics[width=\columnwidth]{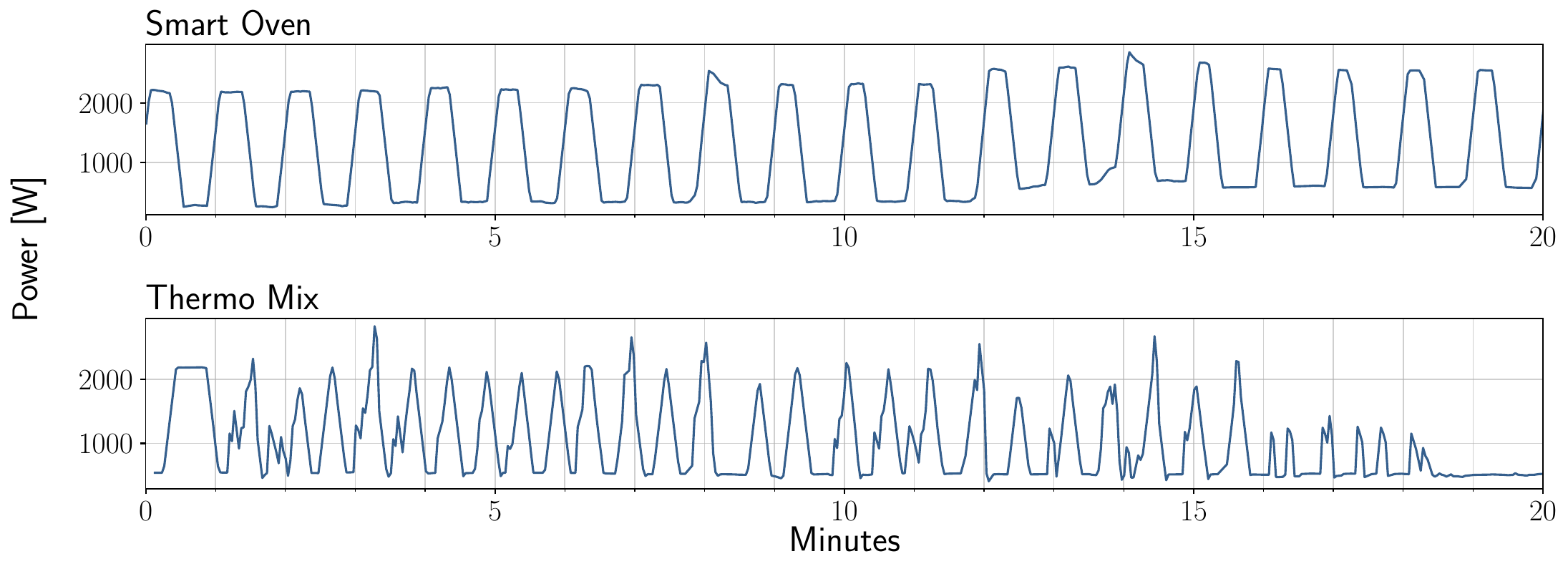}
    \caption{Power time series for two smart devices, a smart baking oven (top) and a Thermomix (bottom).}
    \label{fig:smart_devices}
\end{figure}

With the ongoing “smartification” of energy systems, an increasing number of household and end-user devices are expected to be operated and controlled by internal clocks synchronized via protocols such as NTP. Empirical evidence supports this trend: for example, we observe that a smart baking oven, whose clock is synchronized via the Internet, exhibits a distinct $1 \, \si{min}$ pattern in its power consumption (cf.~Fig.~\ref{fig:smart_devices}) if and only if the oven is operated by setting a timer. The oven’s heating process is triggered at the start of every minute, with the heating demand determining the intensity of the power draw. This results in a stable, recurring $1 \, \si{min}$ power pattern. A Thermomix also displays recurring power peaks; however, its operation does not follow a strict one-minute cycle. These observations highlight the need for further investigation into the behaviour and control of smart devices, as their collective effects may pose additional challenges to frequency stability.

\section{Experiments with computer and ideal Voltage source}

We carry out an experiment to test whether an appliance-level battery in a commercial laptop can exhibit similar patterns in the real power demand as the UPSs. Laptops connected to the Internet generally synchronize their internal clocks via NTP. Hence, it is a priori possible that certain processes start at the beginning of each minute and thus lead to a recurrent variation of the power demand.

\begin{figure}[tb]
    \centering
    \includegraphics[width=0.8\columnwidth]{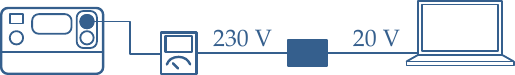}
    \caption{Schematic representation of the setup of the experiment with ideal voltage source.}
    \label{fig:laptop_test}
\end{figure}

To avoid any influence of the external grid, an external battery has been used as ideal voltage source to test the behaviour of the battery of a laptop. Before proceeding with the tests, it has been verified that the battery utilized as ideal voltage source generated current profiles without the peaks observed in the previous section.

\begin{figure}[tb]
    \centering
    \includegraphics[width=.9\columnwidth]{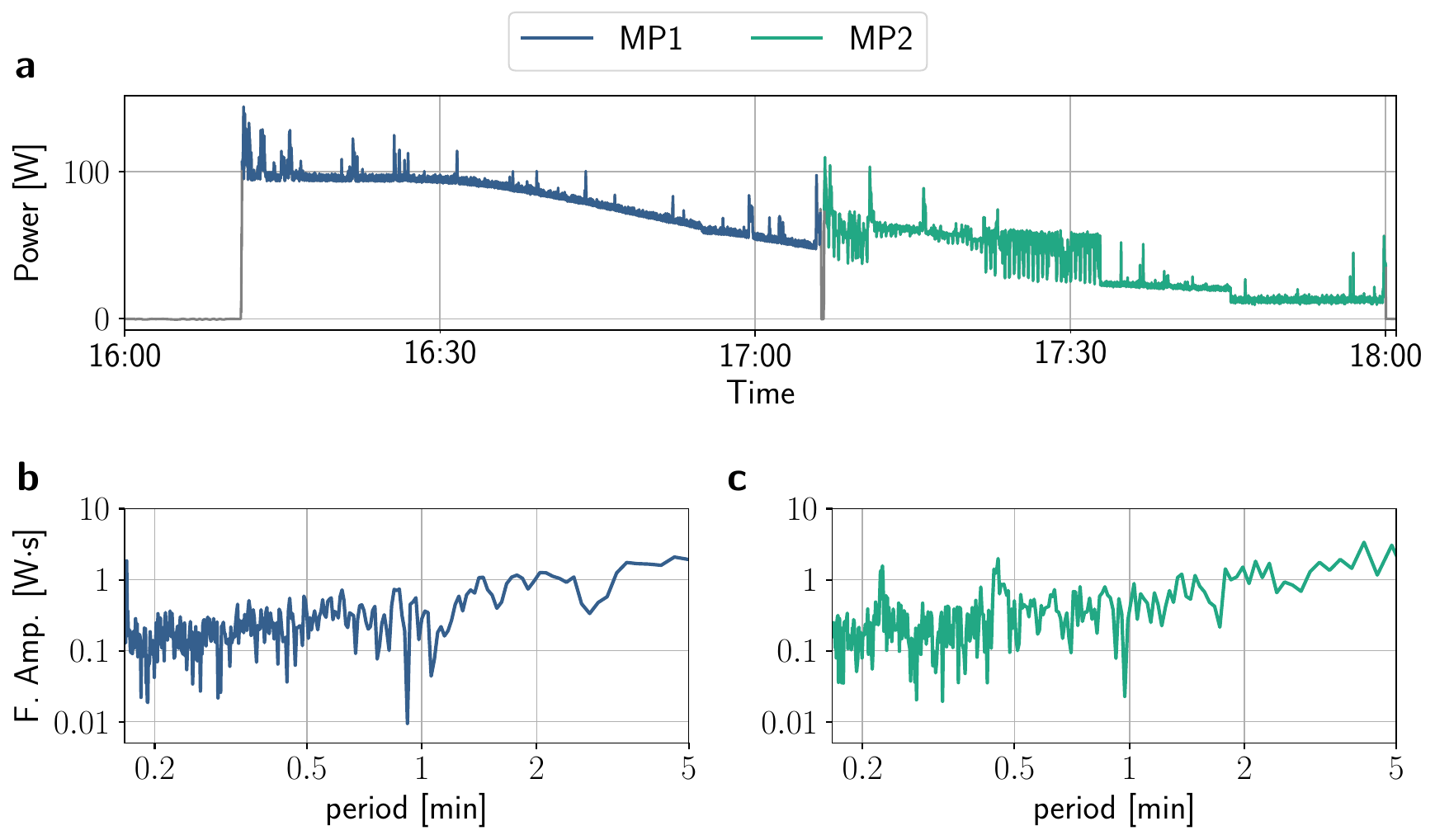}
    \caption{Active Power Experiment Analysis. 
    \textbf{a} Single phase active power time series. Colors indicate the two measurement periods MP1 and MP2 corresponding to the two modes: Youtube video streaming (blue) and idle (green). 
    \textbf{b,c,} Fourier Spectra for both measurement periods.}
    \label{fig:laptop_active_analysis}
\end{figure}

\begin{figure}[tb]
    \centering
    \includegraphics[width=.9\columnwidth]{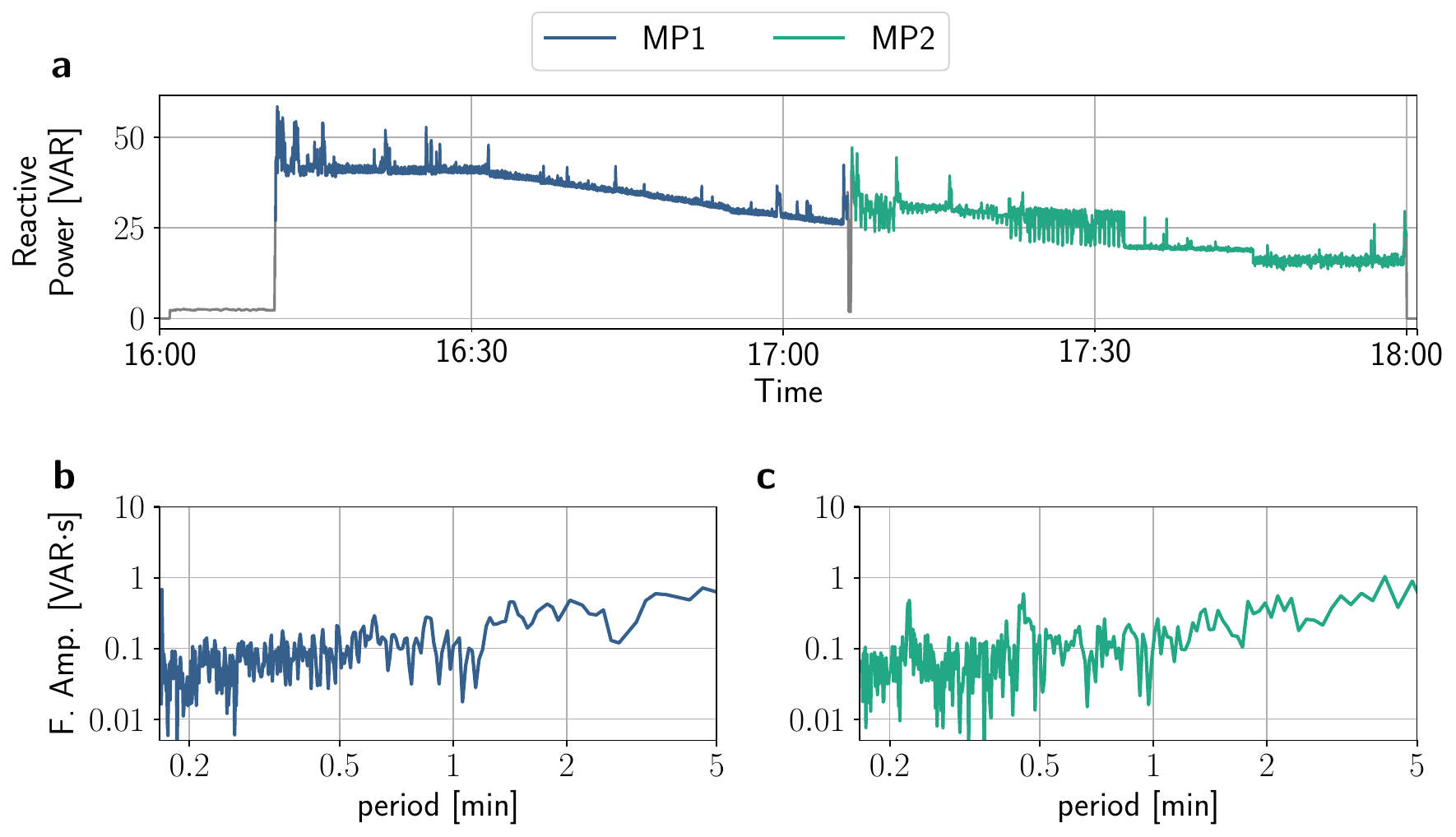}
    \caption{Reactive Power Experiment Analysis. 
    \textbf{a} Single phase reactive power time series.
    Colors indicate the two measurement periods MP1 and MP2 corresponding to the two modes: Youtube video streaming (blue) and idle (green). 
    \textbf{b,c,} Fourier Spectra for both measurement periods.}
    \label{fig:laptop_reactive_analysis}
\end{figure}

As described in Fig.~\ref{fig:laptop_test}, the PQ meter has been used to measure single phase current and voltage between the ideal voltage source and the laptop power supply.  
Figure~\ref{fig:laptop_active_analysis} and Fig.~\ref{fig:laptop_reactive_analysis} show the resulting time series and their relative Fourier spectra for active and reactive power respectively, where colors blue and green represent two measurement periods with different programs running on the Laptop. In the first measurement  period (blue), the Laptop streams and displays videos from YouTube. In the measurement period, the Laptop is idle.

These figures indicate that the low-power battery of the laptop did not generate significant patterns in the active power demand similar to those observed in Fig.~\ref{fig:campus_active_fourier}, thereby suggesting that it would not have any effects on the electrical grid when connected to it.

\section{Processed Frequency Time Series}
\label{sec:data-missing}

\begin{figure}[tb]
    \centering
    \includegraphics[width=\columnwidth]{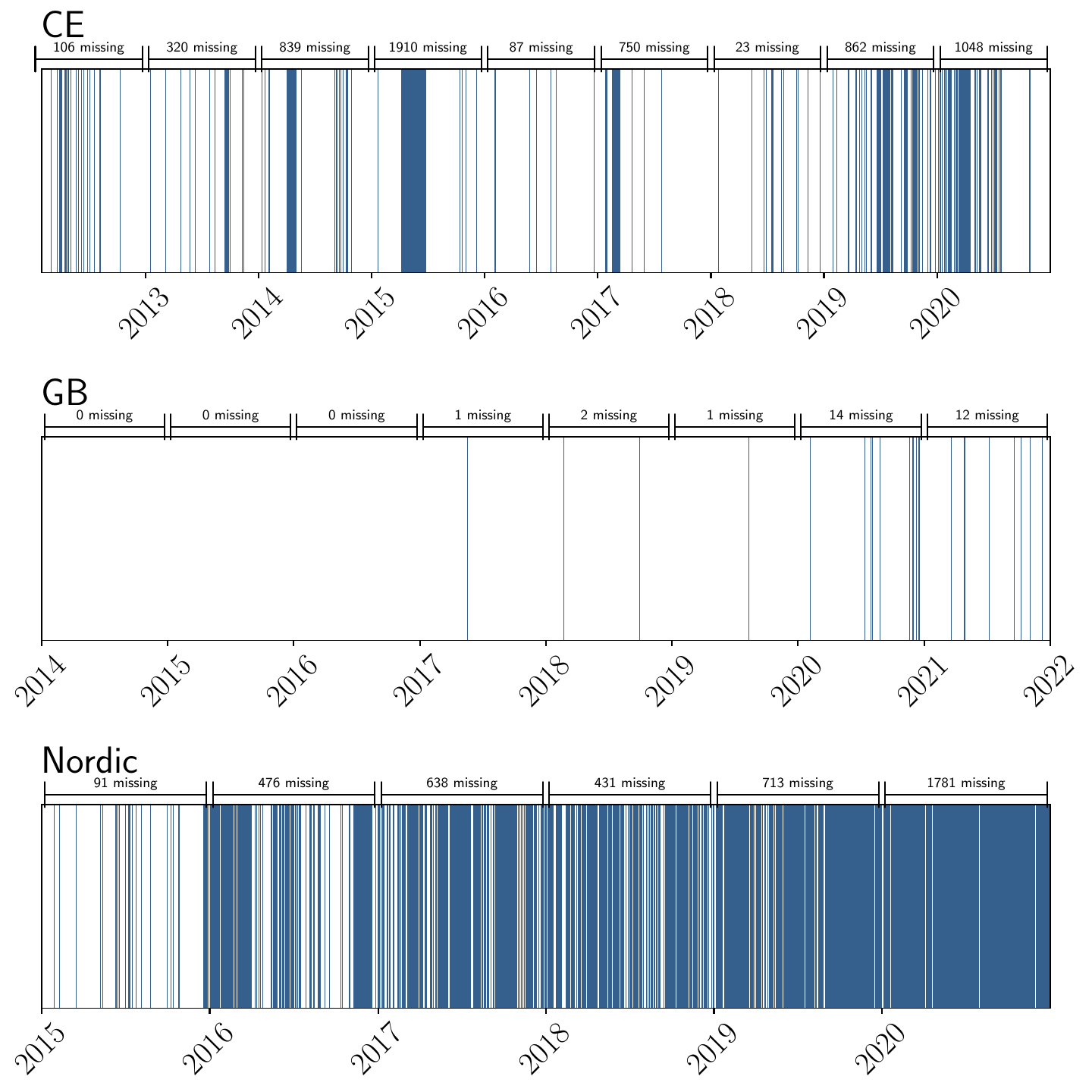}
    \caption{Missing full hours after pre-processing the grid frequency data.}
    \label{fig:missing_freq_data}
\end{figure}

\begin{figure}[tb]
    \centering
    \includegraphics[width=\columnwidth]{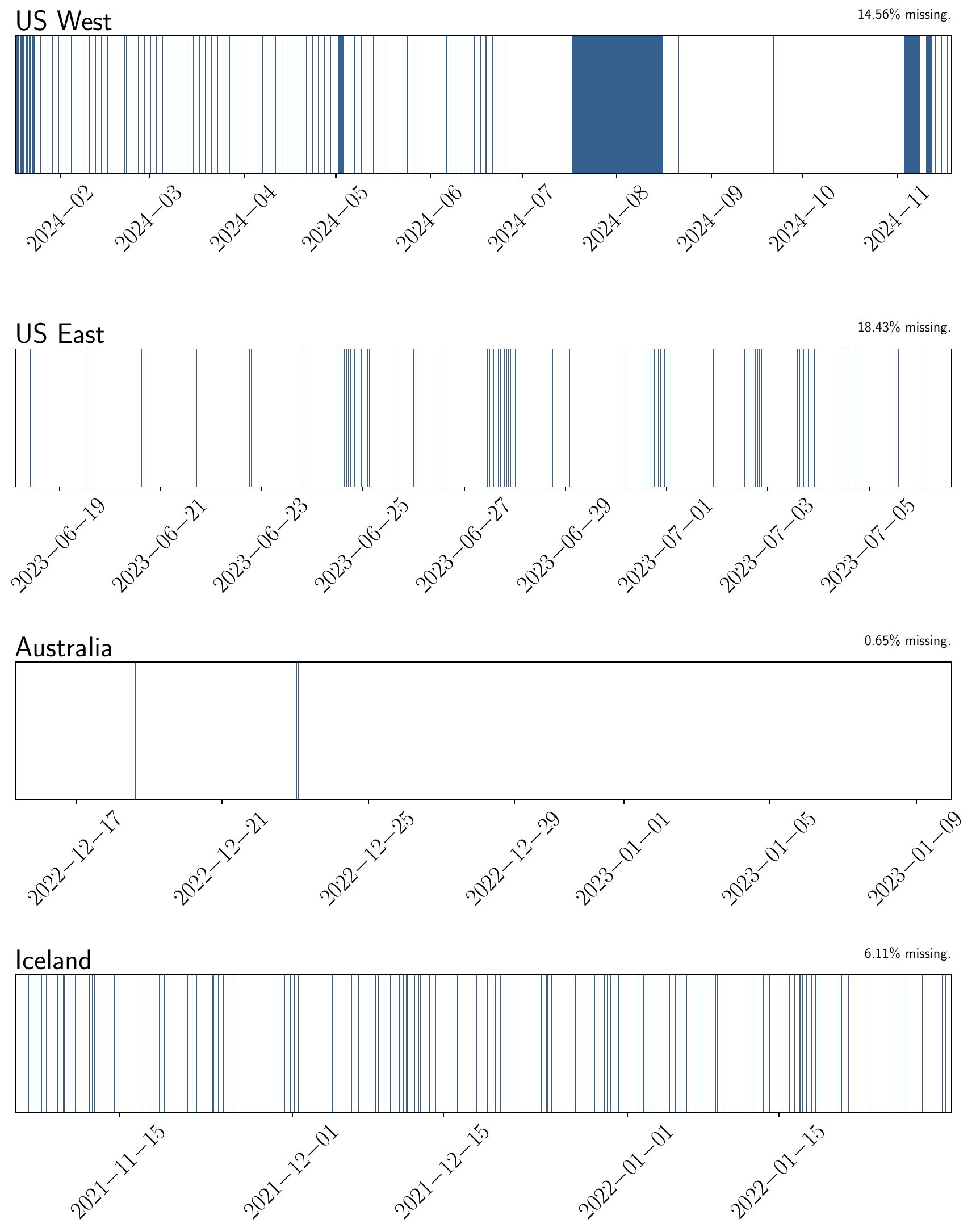}
    \caption{Missing full hours after pre-processing the grid frequency data.}
    \label{fig:missing_freq_data_other_grids_1}
\end{figure}

\begin{figure}[tb]
    \centering
    \includegraphics[width=\columnwidth]{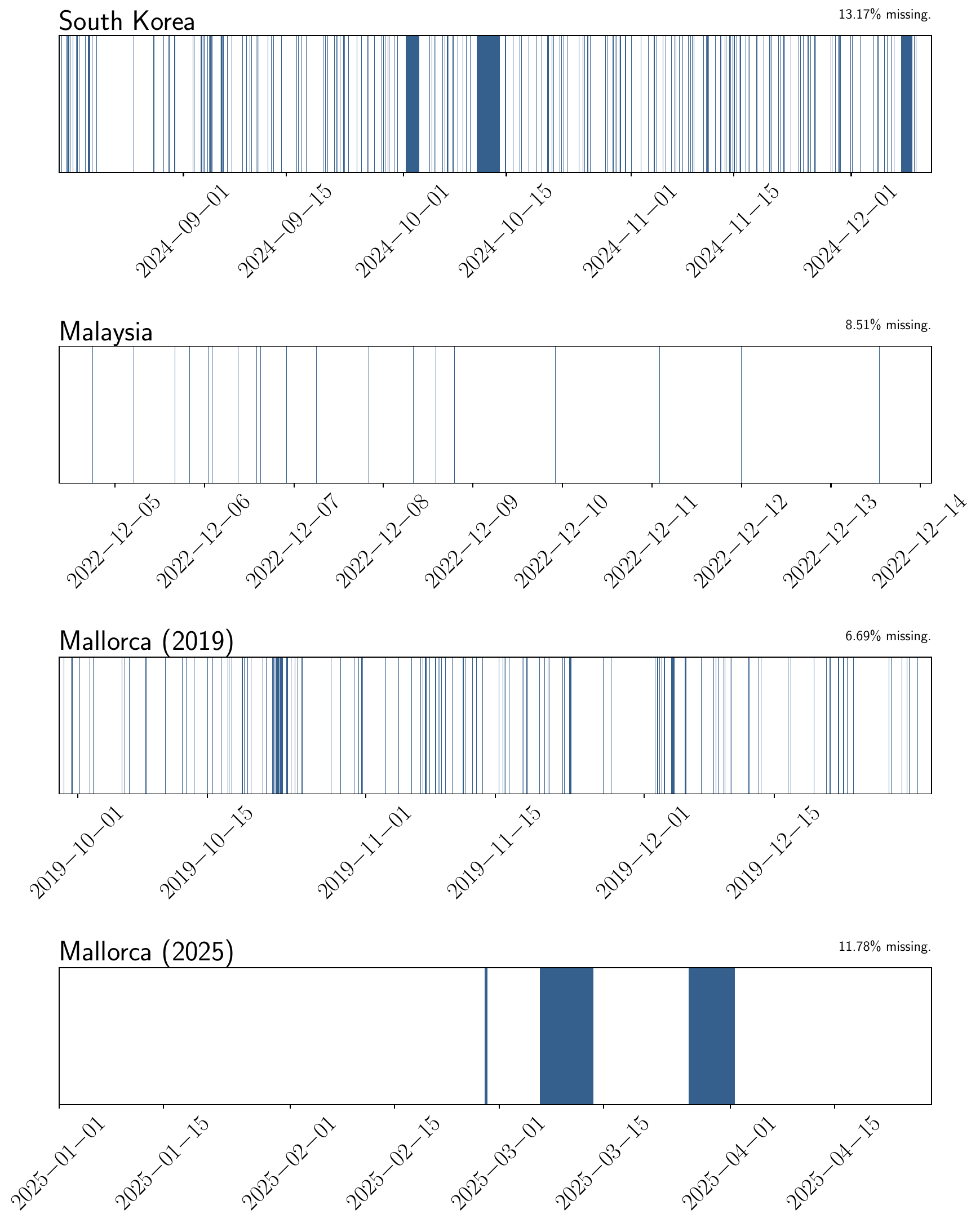}
    \caption{Missing full hours after pre-processing the grid frequency data.
    }
    \label{fig:missing_freq_data_other_grids_2}
\end{figure}

Here, we provide further details about the frequency measurement data. We analyze frequency time series from  three major European synchronous areas provided by the Transmission system operators: the Continental Europe Synchronous Area~\cite{TransnetBW_data}, the Nordic Nordic synchronous area~\cite{Fingrid_data} and the British synchronous area ~\cite{NationalGrid_data}. The measurement periods and the amount of missing data are summarized in Fig.~\ref{fig:missing_freq_data}. 
Frequency time series for other grids are measured on the distribution grid level using the measurement device described in Ref.~\cite{jumar2020power}.The measurement periods and the amount of missing data are summarized in Fig.~\ref{fig:missing_freq_data_other_grids_1} and Fig.~\ref{fig:missing_freq_data_other_grids_2}.

\end{document}